\documentclass[11pt,twoside]{article}
\newcommand{\slsh}[1]{{\not \! #1}}
\newcommand{\be}{\begin{equation}}
\newcommand{\ee}{\end{equation}}
\newcommand{\bea}{\begin{eqnarray}}
\newcommand{\eea}{\end{eqnarray}}
\newcommand{\M}{{\cal M}}

\newcommand{\nn}{\nonumber}
\newcommand{\beas}{\begin{eqnarray*}}
\newcommand{\eeas}{\end{eqnarray*}}
\usepackage{amssymb}
\usepackage{axodraw}
\usepackage{epsfig}

\begin{document}
\title{Gauge Symmetry and its Implications for the Schwinger-Dyson
Equations}

\author{Adnan Bashir and Alfredo Raya}

\date{}

\maketitle
\begin{abstract}

Gauge theories have been a corner stone of the description of the world
at the level of fundamental particles. The Lagrangian or the action
describing the corresponding interactions  is invariant under certain
gauge transformations and contains the dependence on a covariant
gauge parameter. This symmetry is reflected in  terms of the
Ward-Green-Takahashi (or the Slavnov-Taylor) identities (WGTI) which relate various
Green functions among each other, and the Landau-Khalatnikov-Fradkin
transformations  (LKFT) which relate a Green function in a particular gauge to it in
an arbitrary covariant one. As an outcome, all physical observables should be independent 
of the choice of gauge. The most systematic scheme to
solve quantum field theories (QFT) is perturbation theory where the above-mentioned 
identities are satisfied at every order of approximation. This feature is exploited highly
usefully as a verification of the results obtained after time- and effort-consuming exercises.

As it stands, not all natural phenomena are realized in the perturbative
regime of QFTs. Therefore, one is inevitably led to make efforts to solve
these theories in a non perturbative fashion. A natural starting point for
such studies in continuum are Schwinger-Dyson Equations (SDEs). One
of the most undesirable features associated with their non-perturbative
truncations is the loss of highly and rightly esteemed gauge invariance
if special care is not taken. In this work, we present a summary of the modern approach to 
recover gauge invariance in a systematic manner invoking not only WGTI and LKFT but also 
requiring the truncation schemes to  match onto the blindly
trustable perturbative ones at every order of approximation. It is a highly
difficult task but not impossible.
\end{abstract}

\section{Introduction}

In the spectacular realm of quantum fields, gauge 
theories stand out as they lie at the heart of our quest
to unveil the nature of hidden and mysterious forces which
orchestrate the dance of events in the backdrop of the flow of 
time. Whether it be electromagnetic forces which make a
close contact with direct human experience, strong forces
which hold the nuclei intact or weak forces which are manifest
in radioactive decays of elements, gauge theories, namely,
 quantum electrodynamics (QED), chromodynamics (QCD)
and weak dynamics (QWD), seem to provide their accurate description. 
Among these, quantum electrodynamics
enjoys a special place as the quantitative precision of its 
predictions in the perturbative domain is outstandingly high. 
Electromagnetic and weak forces can be wedded 
harmonically into the celebrated Standard Model of particle physics
put forward by Salam, Weinberg and Glashow in late 60's. Its
tremendous success in collating experimental results has frustrated the
scientific community for years which is seeking
physics beyond it. Keeping gravity aloof, in simple extensions of the 
standard model, QCD can also be potentially unified with the other two.

The infallible proof of the  harmonic co-existence of experiment and theory 
for the above-mentioned gauge theories exists neatly only in the perturbative
scheme. When and if the coupling constants involved cease to be small enough to
be treated as an expansion parameter, theoretical
predictions are increasingly hard to make. In some cases, meaningful
predictions of this kind do not even exist. Dynamical chiral symmetry
breaking (DCSB), confinement and the problem of bound states are some examples of
the phenomena which are non perturbative in nature, foreign to
an otherwise wide domain of perturbation theory. It is for this reason
that a true understanding of these phenomena has so far illuded us.
Among several other methods including lattice approach, 
Schwinger-Dyson equations (SDEs),~\cite{esd1,esd2}, provide a natural platform for continuum 
studies of non perturbative phenomena. Although the above-mentioned 
problems of this kind are of direct phenomenological relevance in 
the realm of QCD,~\cite{MRP,HRW,AS,RS,RW} the corresponding SDEs are very complicated and
it becomes relatively more difficult to gain an understanding of its
theoretical intricacies. Within the scope of the present review, 
we concentrate only on the study of DCSB in QED and the problems associated with the 
loss of gauge invariance (GI).

We start with a brief introduction to the Schwinger-Dyson equations to 
set the scene and introduce the notation.  In section 3, 
we recall dynamical generation of fermion masses under the simplest of
approximations. These approximations result in the loss of gauge invariance.
In section 4, we discuss in detail possible sources of this
problem. In next sections, we take up the developments
so far in the resolution of these issues. At the end, we present 
our conclusions.

\section{Schwinger-Dyson Equations}

These equations are an infinite set of coupled integral relations among the 
Green functions of the theory, and form equations of motion of the 
corresponding quantum field theory (QFT)~\cite{esd1,esd2}. The two-point Green 
functions are related to the three-point functions, the three-point functions 
are related to the four-point functions and so on. In perturbation theory, 
fermion propagator can be expressed as in Figure~1.  
\newpage

\vspace*{1cm}
\begin{center}
\SetScaledOffset(-60,0)
\begin{picture}(300,200)(0,0)
\SetScale{0.5}
\ArrowLine(50,450)(150,450)
\CCirc(100,450){5}{}{}
\ArrowLine(200,450)(300,450)
\Vertex(370,450){1} \Vertex(430,450){1}
\ArrowLine(350,450)(450,450)
\PhotonArc(400,450)(30,0,180){4}{8.5}
\PText(175,452)(0)[]{=}
\PText(325,452)(0)[]{+}
\Vertex(220,350){1} \Vertex(280,350){1}
\Vertex(235,350){1} \Vertex(265,350){1}
\ArrowLine(200,350)(300,350)
\PhotonArc(250,350)(30,0,180){4}{8.5}
\PhotonArc(250,350)(15,0,180){3}{6.5}
\ArrowLine(320,350)(500,350)
\PhotonArc(370,350)(30,0,180){4}{8.5}
\PhotonArc(450,350)(30,0,180){4}{8.5}
\PText(175,352)(0)[]{+}
\PText(312,352)(0)[]{+}
\PText(512,352)(0)[]{+}
\Vertex(525,352){2}
\Vertex(535,352){2}
\Vertex(545,352){2}
\Vertex(370,250){1} \Vertex(430,250){1}
\Vertex(390,280){1} \Vertex(410,280){1}
\ArrowLine(350,250)(450,250)
\PhotonArc(400,250)(30,0,70){4}{4.5}
\PhotonArc(400,250)(30,110,180){4}{4.5}
\ArrowArcn(400,280)(10,180,0)
\ArrowArcn(400,280)(10,0,180)
\Photon(400,290)(400,270){2}{4}
\Vertex(280,250){1} \Vertex(220,250){1}
\Vertex(240,280){1} \Vertex(260,280){1}
\Vertex(250,290){1} \Vertex(250,270){1}
\ArrowLine(200,250)(300,250)
\PhotonArc(250,250)(30,0,70){4}{4.5}
\PhotonArc(250,250)(30,110,180){4}{4.5}
\CArc(250,280)(10,0,360)
\PText(175,252)(0)[]{+}
\PText(325,252)(0)[]{+}
\PText(512,252)(0)[]{+}
\Vertex(525,252){2}
\Vertex(535,252){2}
\Vertex(545,252){2}
\Vertex(370,150){1} \Vertex(430,150){1}
\Vertex(400,150){1} \Vertex(460,150){1}
\ArrowLine(350,150)(480,150)
\PhotonArc(400,150)(30,0,180){4}{8.5}
\PhotonArc(430,150)(30,180,0){4}{8.5}
\PhotonArc(430,150)(15,180,0){3}{6.5}
\Vertex(280,150){1} \Vertex(220,150){1}
\Vertex(250,150){1} \Vertex(310,150){1}
\Vertex(265,150){1} \Vertex(295,150){1}
\ArrowLine(200,150)(330,150)
\PhotonArc(250,150)(30,0,180){4}{8.5}
\PhotonArc(280,150)(30,180,0){4}{8.5}
\PText(190,152)(0)[]{+}
\PText(340,152)(0)[]{+}
\PText(512,152)(0)[]{+}
\Vertex(525,152){2}
\Vertex(535,152){2}
\Vertex(545,152){2}
\end{picture}  \\
\vspace{-30pt}
{\sl Figure 1. Perturbative expansion of the fermion propagator.}
\end{center}
The left hand side corresponds to the full two-point
function (the blob over the propagator indicates so), whilst on the right
hand side, we find in the first place the bare
propagator, which describes a fermion propagating with no
perturbative corrections, followed by an infinite set of diagrams for three 
different types of perturbative corrections: to the fermion propagator  
itself, to the boson propagator and to the fermion-boson vertex.   
To be able to perform this infinite sum it is convenient to define 
$\Sigma(p)$ as~:
\be
\Sigma(p) \quad \equiv\qquad\begin{picture}(50,15)(0,0) 
\ArrowLine(0,0)(50,0) 
\PhotonArc(25,0)(12,0,180){3}{4.5}
\CCirc(25,0){3}{}{}
\CCirc(37,0){3}{}{}
\CCirc(25,12){3}{}{}
\end{picture}  \;. 
\ee
This is called the self energy and involves all the
corrections (as indicated by the blob) to the fermion propagator, to the 
boson propagator
and to the fermion-boson vertex.  In terms of self energy,
perturbative expansion of the fermion propagator is despicted in
Figure~2, 
\begin{center}
\begin{picture}(300,30)(0,0)
\ArrowLine(0,0)(50,0)
\ArrowLine(70,0)(120,0)
\ArrowLine(140,0)(190,0)
\Line(210,0)(295,0)
\PhotonArc(165,0)(12,0,180){3}{4.5}
\PhotonArc(230,0)(12,0,180){3}{4.5}
\PhotonArc(275,0)(12,0,180){3}{4.5}
\CCirc(25,0){3}{}{}
\CCirc(165,0){3}{}{}
\CCirc(177,0){3}{}{}
\CCirc(165,12){3}{}{}
\CCirc(230,0){3}{}{}
\CCirc(242,0){3}{}{}
\CCirc(230,12){3}{}{}
\CCirc(275,0){3}{}{}
\CCirc(275,12){3}{}{}
\CCirc(287,0){3}{}{}
\Text(60,0)[]{=}
\Text(130,0)[]{+}
\Text(200,0)[]{+}
\Text(310,0)[]{+$\ldots$} 
\end{picture} \\
\vspace{20pt}
{\sl Figure 2. Perturbative expansion of the fermion propagator in terms of 
the self energy.}
\end{center}
and corresponds to the sum
\be
S_F(p)=S_F^0(p)+S_F^0(p)\Sigma(p) S_F^0(p)+S_F^0 \Sigma(p)
S_F^0(p)\Sigma (p)S_F^0(p)+\ldots   \;.
\ee
Factorizing $S_F^0(p)$, the remaining geometric series yields
\be
S_F(p)= \frac{S_F^0(p)}{1-\Sigma(p)S_F^0(p)}\;.\label{geom1}
\ee
Equivalently, leaving the first term apart and factorizing $S_F^0
\Sigma(p) S_F^0(p)$, the remaining terms form once again a
geometric series, which after summation yields
\be
S_F(p)=S_F^0(p)+S_F^0(p)\Sigma(p)
\frac{S_F^0(p)}{1-\Sigma(p)S_F^0(p)}\;.\label{geom2}
\ee
Comparing (\ref{geom1}) and (\ref{geom2}) we obtain
\be
S_F(p)=S_F^0(p)+S_F^0(p)\Sigma(p)S_F(p)\;.
\ee
This is the SDE for the fermion
propagator.  It is customary to write and study the SDE
for the inverse fermion propagator instead, which is
\be
S_F^{-1}(p)=S_F^{0-1}(p)-\Sigma(p)  \;,\label{SDEfp}
\ee
and corresponds to the diagram
\vspace*{-10pt}
\begin{center}
\begin{picture}(500,65)(0,0)
\SetScale{0.7}
\ArrowLine(50,50)(150,50)
\CCirc(100,50){5}{}{}
\PText(145,60)(0)[]{-1}
\ArrowLine(200,50)(300,50)
\PText(295,60)(0)[]{-1}
\Vertex(430,50){1} \Vertex(370,50){1}
\ArrowLine(350,50)(450,50)
\PhotonArc(400,50)(30,0,180){4}{8.5}
\CCirc(400,80){5}{}{}
\CCirc(430,50){5}{}{}
\CCirc(400,50){5}{}{}
\PText(175,52)(0)[]{=}
\PText(325,52)(0)[]{-}
\PText(465,55)(5)[]{.}
\end{picture}    \\
\vspace{-20pt}
{\sl Figure 3. The SDE for the (inverse) fermion propagator.}
\end{center}
Of course, the full Green functions involved in the self energy
obey their own SDE.  The one for the
boson propagator is shown in Figure~4
\vspace*{-1cm}
\begin{center}
\begin{picture}(500,100)(0,0)
\SetScale{0.7}
\Photon(50,50)(150,50){4}{10}
\PText(145,60)(0)[]{-1}
\CCirc(100,50){5}{}{}
\Photon(200,50)(300,50){4}{10}
\PText(295,60)(0)[]{-1}
\Vertex(370,50){1}
\ArrowArcn(400,50)(30,180,0)
\ArrowArcn(400,50)(30,0,180)
\Photon(350,50)(370,50){4}{2}
\Photon(430,50)(450,50){4}{2}
\CCirc(430,50){5}{}{}
\CCirc(400,80){5}{}{}
\CCirc(400,20){5}{}{}
\PText(175,52)(0)[]{=}
\PText(325,52)(0)[]{-}
\end{picture} \\
{\sl Figure 4. The SDE for the boson propagator.}
\end{center}
and the one for the fermion-boson vertex is depicted in
Figure~5. 
\begin{center}
\begin{picture}(500,60)(0,0)
\SetScale{0.7}
\Line(75,50)(125,75)
\Line(75,50)(125,25)
\Vertex(75,50){1}
\Photon(25,50)(75,50){-3}{4}
\CCirc(75,50){5}{}{}
\Line(225,50)(275,75)
\Line(225,50)(275,25)
\Vertex(225,50){1}
\Photon(175,50)(225,50){-3}{4}
\Line(425,50)(475,75)
\Line(425,50)(475,25)
\Vertex(375,50){1}
\Photon(325,50)(375,50){-3}{4}
\CArc(400,50)(25,0,360)
\CCirc(425,50){5}{}{}
\CCirc(400,75){5}{}{}
\CCirc(400,25){5}{}{}
\CCirc(375,50){5}{}{}
\PText(150,52)(0)[]{=}
\PText(300,52)(0)[]{-}
\PText(500,52)(0)[]{.}
\end{picture}  \\
\vspace{0pt}
{\sl Figure 5. The SDE for the fermion-boson vertex.}
\end{center}
We notice that the two-point functions are coupled
to each other and both are coupled to a three-point function.  The
vertex, in turn, is coupled to two-point,
three-point and four-point Green functions 
and so on, leading
therefore, to an infinite set of coupled equations which form the field
equations of the theory.
Because of the Lorentz structure of Green functions the
fermion propagator can be written in terms of two scalar
functions. We prefer to write it as
\be
S_F(p) = \frac{F(p)}{\slsh{p}-\M(p)}\;.\label{FPdef}
\ee
$F(p)$ is generally referred to as the wavefunction renormalization and
$\M(p)$ as the mass function.  Pole of the propagator
corresponds to the physical mass of the fermion.
The gauge boson propagator can in general be written as 
\be
\Delta^{\mu\nu}(q)= \frac{ {\cal G} (q)}{q^2}\left(g^{\mu\nu}-\frac{q^\mu
q^\nu}{q^2} \right)+\xi \frac{q^\mu q^\nu}{q^4}\;. \label{PPdef}
\ee
${\cal G}(q)$ is called the gauge boson wavefunction renormalization and
$\xi$ is the usual covariant gauge parameter.
The fermion-boson vertex $\Gamma^\mu(k,p)$ can be expressed in
terms of twelve Lo\-rentz structures as 
\be
\Gamma^\mu(k,p)= \sum_{i=1}^{12} v_i(k,p)V_i^\mu\;,
\ee
where
\be
\begin{array}{ccc} V_1^\mu = \gamma^\mu, & V_2^\mu =k^\mu, &
V_3^\mu = p^\mu,\\ &&\\
V_4^\mu = \slsh{k}\gamma^\mu, & V_5^\mu= \slsh{k}
k^\mu, & V_6^\mu= \slsh{k}p^\mu,  \\ &&\\
V_7^\mu = \slsh{p}\gamma^\mu, & V_8^\mu= \slsh{p}
k^\mu, & V_9^\mu= \slsh{p}p^\mu,\\ && \\
V_{10}^\mu = \slsh{k}\slsh{p}\gamma^\mu, & V_{11}^\mu= \slsh{k}\slsh{p}
k^\mu, & V_{12}^\mu= \slsh{k}\slsh{p}p^\mu.\\ \\
\end{array}   \label{FVdef}
\ee
From here, we can infer that
the higher the order of the Green function, the more complicated is its
structure and the corresponding SDE. In order
to truncate the infinite tower of SDEs, special care should be
taken to preserve gauge invariance and its implications.

If we want to truncate the infinite tower of SDEs at the level of two-point 
Green functions, a popular way is to make an \emph{ansatz} for the 
fermion-boson interaction which preserves key features of QED, such as the 
gauge covariance of the Green functions and 
gauge invariance of associated physical observables. 
In the following, we shall review how some of the consequences of
gauge invariance could put constraints on the non perturbative
truncations of the SDE studies, namely the
Ward-Green-Takahashi identities (WGTI) and the
Landau-Khalatnikov-Fradkin transformations (LKFT),  and 
how perturbation theory, being the
only known scheme where gauge invariance is satisfied at every
order of approximation, could play a guiding role in this context.

\section{Dynamical Chiral Symmetry Breaking}

As we mentioned before, SDEs are an ideal framework to study non perturbative 
phenomena in gauge theories, such as  dynamical breaking 
of chiral symmetry, which we shall study in QED. In order to proceed, we 
must recall that eq.~(\ref{SDEfp}) can be cast in arbitrary dimensions in the 
following way
\be
S_F^{-1}(p)= S_F^{0\,-1}(p)-4 \pi i \alpha  \int d^dk \gamma^\mu S_F(k) 
\Gamma^\nu(k,p) \Delta_{\mu\nu}(q)\;. \label{SDEfpd}
\ee
To be able to solve it, we must know the explicit form of the 
gauge boson propagator 
and the fermion-boson vertex. However, we can simplify this equation in the 
so-called rainbow approximation by neglecting fermion loops (i.e., setting 
${\cal G} (q)=1$) and considering $\Gamma^\mu(k,p)=\gamma^\mu$. In such case, 
we are led to the expression
\be
S_F^{-1}(p)= S_F^{0\,-1}(p)-4 \pi i\alpha  \int d^dk \gamma^\mu S_F(k) 
\gamma^\nu \Delta^0_{\mu\nu}(q)\;. \label{SDEfprainbow}
\ee
Diagramatically it corresponds to 
\vspace*{-1cm}
\begin{center}
\begin{picture}(500,100)(0,0)
\SetScale{0.7}
\ArrowLine(50,50)(150,50)
\CCirc(100,50){5}{}{}
\PText(145,60)(0)[]{-1}
\ArrowLine(200,50)(300,50)
\PText(295,60)(0)[]{-1}
\Vertex(430,50){1} \Vertex(370,50){1}
\ArrowLine(350,50)(450,50)
\PhotonArc(400,50)(30,0,180){4}{8.5}
\CCirc(400,50){5}{}{}
\PText(175,52)(0)[]{=}
\PText(325,52)(0)[]{-}  
\end{picture}
{\sl Figure 6. The SDE for the fermion propagator in rainbow approximation}
\end{center}
and deserves its name because it correspond to a fermion propagator of the 
 form 
\begin{center}
\begin{picture}(300,120)(0,0)
\ArrowLine(0,60)(75,60)
\Vertex(37,60){3}
\ArrowLine(100,60)(175,60)
\ArrowLine(200,60)(275,60)
\PhotonArc(237,60)(25,0,180){4}{8.5}
\PText(87,62)(0)[]{=}
\PText(187,62)(0)[]{+}
\ArrowLine(100,0)(175,0)
\PhotonArc(137,0)(25,0,180){4}{8.5}
\PhotonArc(137,0)(12,0,180){3}{8.5}
\ArrowLine(200,0)(275,0)
\PhotonArc(237,0)(25,0,180){4}{8.5}
\PhotonArc(237,0)(12,0,180){3}{8.5}
\PhotonArc(237,0)(6,0,180){2}{8.5}
\PText(87,2)(0)[]{+}
\PText(187,2)(0)[]{+}
\PText(287,2)(0)[]{+...}
\end{picture} .\\ 
\vspace{1cm}
{\sl Figure 7. Fermion propagator in rainbow approximation.}
\end{center}
This is probably the simplest approximation in which the DCSB can be studied. 
We shall recall the solution of eq.~(\ref{SDEfprainbow}) in two cases: $d=4$ and 
$d=3$. In the former case, an extra complication occurs because of the
presence of ultraviolet divergences. The numerical method is the
simplest to employ if one uses the cut-off method to deal with these
infinities. In this case, it could well be that not only the simplifying 
assumptions are the root of gauge invariance, but also the regularization 
procedure. The case of QED3 is neater in the sense that since it is a 
superrenormalizable theory, we do not have to deal with the possible
gauge dependence problems arising from regularization issues. 
Instead, we can focus entirely on the assumptions made in the truncation of 
SDEs and how gauge symmetry and its implications can help us solve the 
problem  of gauge invariance.

\subsection{DCSB in QED4}

Eq.~(\ref{SDEfprainbow}) is a matrix equation which can be rearranged into 
a system of two coupled equations for $F(p)$ and $\M (p)$ on 
multiplying it by $\slsh{p}$ and 1, respectively, and taking the trace. 
After performing a Wick rotation to Euclidean space~\footnote{Attempts have been made to solve the SDEs in Minkowski space, see for example~\cite{Sauli}.} and setting $d=4$, 
angular integrations can be readily done. Radial integrations thus yield
\bea
\frac{\M(p)}{F(p)} &=& m_0 + \frac{\alpha}{4\pi}(3+\xi) 
\int_0^{\Lambda^2} dk^2 \frac{F(k)\M(k)}{k^2+\M^2(k)} 
\left[ \frac{k^2}{p^2}\theta(p^2-k^2)+\theta(k^2-p^2)\right]\nn\\
\frac{1}{F(p)}&=& 1+\frac{\alpha\xi}{4\pi} \int_0^{\Lambda^2} dk^2  
\frac{F(k)}{k^2+\M^2(k)}\left[ \frac{k^4}{p^4}\theta(p^2-k^2)+
\theta(k^2-p^2)\right]\;,\label{coupledFM4}
\eea
where $\Lambda$ is the ultraviolet cut-off scale for the internal momentum
and $m_0$ is the bare fermion mass.
This system of equations has to be simultaneously solved by numerical techniques 
for every value of the gauge parameter in order to study the gauge dependence of the associated physical observables.
One can  study this system of coupled equations to look for 
dynamically generated mass,~\cite{initial1,initial2,initial3,initial4}. 
We observe that in the Landau gauge 
($\xi=0$), $F(p)=1$ which lead us to solve but one equation:
\be
\M(p) = m_0 + \frac{3\alpha}{4\pi} \left[\frac{1}{p^2}\int_0^{p^2} 
dk^2 \frac{k^2\M(k)}{k^2+\M^2(k^2)} +\int_{p^2}^{\Lambda^2} 
\frac{\M(k^2)}{k^2+\M(k^2)}\right]\;. \label{Mlg4}
\ee
Setting $m_0=0$,
results are shown in Figure~8. One can also  translate eq.~(\ref{Mlg4})  into a 
differential equation 
with the appropriate boundary conditions. On doing so, we find
\be
\frac{d}{dp^2}\left(p^4 \frac{d\M(p)}{dp^2} \right)=
-\frac{3\alpha}{4\pi}\frac{p^2\M(p)}{p^2+\M^2(p)}\;,\label{diffeqM4}
\ee
with the boundary conditions 
\be
\frac{1}{p^2}\left[k^4 \frac{d\M(k)}{dk^2} \right]_{k^2=0}\to 0 
\qquad \mbox{and} \qquad  \left[\M(k)+k^2  \frac{d\M(k)}{dk^2} 
\right]_{k^2=\Lambda^2} \to 0\;.
\ee

\begin{center}
\epsfig{file=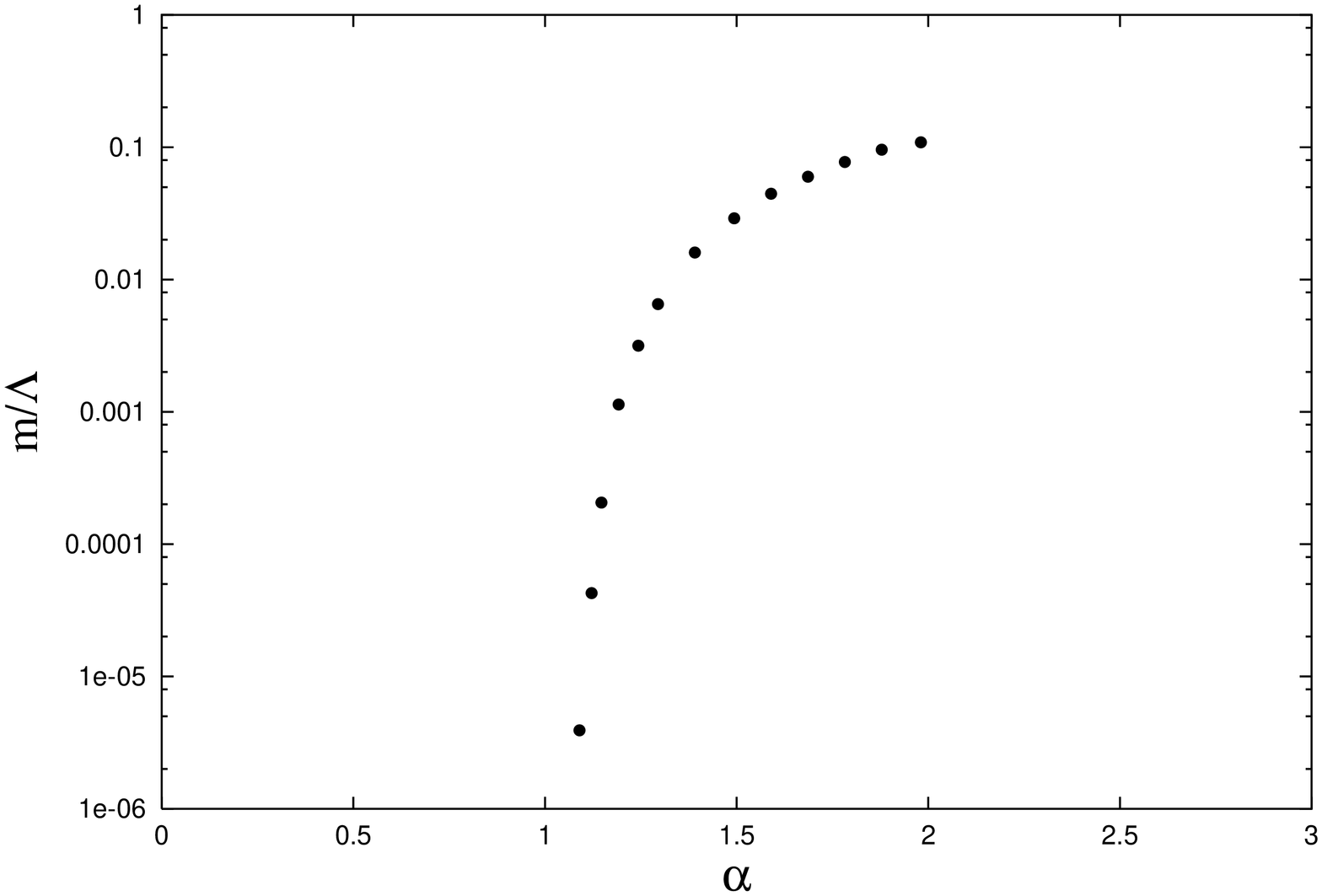,width=0.4\textwidth,angle=-90}
\\
\vspace{0.5cm}
{\sl Figure 8. Dynamically generated mass as a function of the coupling.}
\end{center}

\noindent
In the limit $p^2\gg \M^2(p)$, eq.~(\ref{diffeqM4}) simplifies to
\be
\frac{d}{dp^2}\left(p^4 \frac{d\M(p)}{dp^2} \right)\simeq-
\frac{3\alpha}{4\pi}\M(p)\;,
\ee
which leads us to the following power-law behaviour for the mass function
\be
\M(p)\sim (p^2)^s \qquad \mbox{with} \qquad s(s+1)=-\frac{3\alpha}{4\pi}\;.
\ee
Just like on the numerical graph, we observe that there is a critical value 
$\alpha_c$ of the coupling which distinguishes a pure power-law behaviour 
from an oscillatory behavior for the mass function:
\be
\alpha_c= \frac{\pi}{3}\;.
\ee
This critical number tells us that in order for the fermion masses to be 
generated dynamically, we need a coupling constant above $\pi/3$.
This critical coupling, in principle, is a physical observable.
Unfortunately, as we shall see later, its value varies as a function of the 
gauge  parameter which implies that our truncation procedure is not entirely 
reliable. We shall later come back to this problem and its possible solutions.

\subsection{DCSB in QED3}

Performing a similar exercise in eq.~(\ref{SDEfprainbow}) for $d=3$,  we are 
left with the following system of equations
\bea
\frac{1}{F(p)}&=& 1-\frac{\alpha\xi}{4\pi}\int_0^\infty dk 
\frac{k^2F(k)}{k^2+\M^2(k)}\left[1-\frac{k^2+p^2}{2kp}\ln{\left| 
\frac{k+p}{k-p}\right|} \right]\;,\nn\\
\frac{\M(p)}{F(p)}&=& \frac{\alpha(\xi+2)}{\pi p} \int_0^\infty dk 
\frac{k F(k) \M(k)}{k^2+\M^2(k)}\ln{\left| \frac{k+p}{k-p}\right|}\;.
\eea
Again, in Landau gauge we find $F(p)=1$ and thus we are led to the equation
\be
\M(p)=\frac{2\alpha}{\pi p} \int_0^\infty dk \frac{k 
\M(k)}{k^2+\M^2(k)}\ln{\left| \frac{k+p}{k-p}\right|}\;.\label{Mlg3d}
\ee
Numerical solution of this equation also reveals dynamically generated
mass function. However, unlike QED4, there is no critical value of 
coupling. If masses are generated for one value of coupling, they are
also generated for any other value. 
Again, 
the large-momentum behaviour of 
the mass function can also be inferred by translating this equation into a 
differential equation with the appropriate boundary conditions. For this 
purpose, let us observe that we can make the approximation
\be
\ln{\left| \frac{k+p}{k-p}\right|} \simeq \frac{2p}{k}\theta(k-p)
+\frac{2k}{p}\theta(p-k)
\ee
which is valid for $k\gg p$ as well as for $k\ll p$. Thus, we can rewrite
eq.~(\ref{Mlg3d}) as
\be
\frac{d}{dp}\left[p^3 \frac{d\M(p)}{dp} \right]
+\frac{2}{\pi^2}\M(p)=0 \label{diffeqM3}
\ee
with the boundary conditions
\be
\left[p^3 \frac{d\M(p)}{dp}\right]_{p=0}\to 0 
\qquad \mbox{and} \qquad \M(p)_{p\to\infty}\to 0\;.
\ee
The solution to~(\ref{diffeqM3}) is given by
\be
\M(p)= \frac{4}{ \pi^2 p}
\left[c_1 J_2\left(\sqrt{\frac{8}{\pi^2 p}}\right) + 
c_2 Y_2\left(\sqrt{\frac{8}{\pi^2 p}}\right)\right]\;,
\ee
where $J(x)$ and $Y(x)$ are Bessel functions of the first and second kind, 
respectively. The second boundary condition imposes $c_2=0$. 
As $J(1/\sqrt{p})\to 1/p$ for $p\to \infty$, we conclude that $\M(p)\to 
1/p^2$ in this limit. $c_1$ cannot be determined from the equation and 
boundary conditions. The analysis for different values of the gauge parameter 
reveals an undesirable fact: the chiral condensate $\langle\bar\psi\psi\rangle$, 
which is a relevant physical observable in this case, depends also on the gauge parameter.
The moral is that special care should be taken in the truncation of the tower 
of SDEs to preserve gauge invariance. 
The way to remedy this situation in the quenched case is to make use of 
an 
\emph{ansatz} for the fermion-boson interaction which preserves key features 
of QED. These restrictions and their implementations are detailed in the 
next sections.

\section{Restrictions on the Vertex}

The constraints on the choice of the three-point vertex are as follows~:
\begin{itemize}

\item  {\bf {\underline{Ward Identities:}}} 

Ward and Ward-Green-Takahashi identities, i.e.,
\bea
\Gamma^\mu(p,p) &=& \frac{\partial}{\partial p_\mu} S_F^{-1}(p)\;, \label{WI}
\\  \nn \\ 
q_\mu \Gamma^\mu(k,p) &=& S_F^{-1}(k)-S_F^{-1}(p)             \;, \qquad q=k-p\label{WGTI}
\eea
which relate the fermion propagator to the three-point vertex should be satisfied.

\item   {\bf {\underline{Gauge Independent Physical Observables:}}}

Physical observables related with the fermion propagator such as
the physical mass of the fermion, the condensate, etc. should be independent
of the gauge parameter.

\item   {\bf {\underline{Gauge Independent Regularization Scheme:}}}

  Although it is not a constraint on the vertex and is only a technical
detail, it is important enough to be taken carefully into consideration.
A gauge dependent ultraviolet regulator used carelessly can add additional
and incorrect gauge dependence in the study of Schwinger-Dyson equations.


\item {\bf {\underline{Kinematic Singularities:}}} 

It should be independent of any kinematic singularities such
as the ones when $k^2 \rightarrow p^2$.

\item {\bf {\underline{Transformation Under C, P and T:}}} 

Under the operations of charge conjugation, parity and time reversal, 
it should transform in the same way as the bare vertex $\gamma^{\mu}$.

\item {\bf {\underline{Correct Weak Coupling Limit:}}} 

It should reduce to the perturbative Feynman expansion for the vertex 
in the limit when the coupling is small.

\item {\bf {\underline{Landau-Khalatnikov-Fradkin Transformation Laws:}}}  

        Under a variation of gauge, the fermion propagator and the 
fermion-boson vertex should change in accordance with
 the Landau-Khalatnikov-Fradkin transformations (LKFT). In the position 
space,
the LKFT for the fermion propagator reads~:
\bea
      S(x;\xi) &=&  S(x;0) \; {\rm e}^{-[\Delta_d(0)-\Delta_d(x)]} \;,
\eea
where
\bea
     \Delta_d(x) &=& \frac{\xi e^2}{16 \pi^{d/2}} \; (\mu x)^{4-d}
\; \Gamma \left(\frac{d}{2} -2 \right) \;. \nn
\eea
The corresponding law for the vertex is more complicated in form.

\end{itemize}

\section{The Ward Identities}

As a consequence of the gauge invariance principle of QED, its  
Green  functions  are  related  through Ward identities, 
\cite{W1,G1,T1}. Particularly, the  
fermion  propagator  is related to
the fermion-boson vertex, eq.~(\ref{WGTI}).
This  identity
is non-perturbative. Any choice of the vertex which does 
not satisfy this relation will violate the gauge principle. 
Let us first look at the bare vertex, i.e., $\Gamma^\mu(k,p)=\gamma^{\mu}$.
In this case
\bea
\slsh{q} &=&  \frac{\slsh{k}}{F(k)} - \frac{\slsh{p}}{F(p)} 
- \frac{{\cal M}(k)}{F(k)} + \frac{{\cal M}(p)}{F(p)}   \;.  \nn
\eea
Obviously, this equation cannot be satisfied in every gauge for the
dynamically generated fermion propagator. What about the Landau gauge?
We know that in the Landau gauge $F(k)=F(p)=1$. Moreover, the mass
function is nearly flat below the pole position ${\cal M}^2(p)=p^2$.
One can then conclude that the bare vertex practically satisfies the WGTI 
for the chirally asymmetric fermion propagator in the Landau gauge for
the values of $k^2$ and $p^2$ below the physical (Euclidean) pole mass of
the fermion. Therefore, if this is the physical observable we are interested
to calculate, bare vertex in the Landau gauge should be sufficient as far
as the WGTI is concerned. Will this approximation lead to the correct 
result? The answer is: not necessarily, 
although one may hope to be in its vicinity. The reason is the
following. Imagine we could construct a vertex (which is easy to do in 
practice)
which in fact satisfies the WGTI in every gauge exactly. Now we are
free to do the calculation in any gauge. When we do this exercise, we find
that the result for the physical observables in general varies from gauge 
to gauge. Therefore, having the satisfaction that the bare vertex satisfies the WGTI
in the Landau gauge can not translate into a tangible statement for
the gauge independence of physical observables. However, WGTI is a necessary
condition and it must be satisfied by any vertex constructed.

The structure of the WGTI is such that we can decompose the full vertex
into the following two components, longitudinal and transverse~:
\bea
\Gamma^\mu(k,p)= \Gamma^\mu_{L}(k,p)+\Gamma^\mu_T(k,p)\;,
\eea
where, by definition the transverse vertex $\Gamma^\mu_T(k,p)$ is such that 
$q_\mu \Gamma^\mu_T=0$. 
A possible choice for the longitudinal vertex is,~\cite{AJS,Kondo}~: 
\bea
\Gamma^\mu_L(k,p)= 
\frac{q^\mu}{q^2}[S_F^{-1}(k)-S_F^{-1}(p)]\;. \label{LV1}
\eea
However, this has a singularity for $q^2 \rightarrow 0$. No such singularity
is encountered in perturbation theory for the vertex at the one loop level,
which makes it an unacceptable choice. As the longitudinal vertex is not
unique, several proposals exist in literature,~\cite{BC,Rambiesa,Haeri}.
More recently, it has been a common practice to employ as the longitudinal
vertex the one proposed by Ball and Chiu,~\cite{BC}. This 
\emph{ansatz,} is based upon the Ward identity, eq.~(\ref{WI}).
Substituting in it the expression for the full fermion propagator, we observe that
\bea
\Gamma^\mu(p,p)
&=&\frac{\gamma^\mu}{F(p)}+2p^\mu\slsh{p}\frac{\partial}{\partial p^2}
-2p^\mu\frac{\partial}{\partial p^2}\frac{{\cal M}(p)}{F(p)}\;.
\eea
After we make the symmetric substitutions for the case when $k \neq p$
\bea
\frac{1}{F(p)}&\to&\frac{1}{2}\left[\frac{1}{F(k)}+\frac{1}{F(p)}
\right]\; ,\nn\\
p^\mu &\to& \frac{1}{2}(k^\mu+p^\mu) \; ,\nn\\
\slsh{p}&\to&\frac{1}{2}(\slsh{k}+\slsh{p})\;, \nn\\
\frac{\partial}{\partial p^2}\frac{1}{F(p)}&\to&
\frac{1}{k^2-p^2}\left[ \frac{1}{F(k)}-\frac{1}{F(p)}\right] \; ,\nn\\
\frac{\partial}{\partial p^2}\frac{{\cal M}(p)}{F(p)}&\to&
\frac{1}{k^2-p^2}\left[ \frac{{\cal M}(k)}{F(k)}
-\frac{{\cal M}(p)}{F(p)}\right]\;,
\eea
Ball and Chiu defined the longitudinal vertex as,~\cite{BC},~:
\bea
\Gamma^{\mu}_{BC}&=&\frac{\gamma^{\mu}}{2}
\left[ \frac{1}{F(k)}+\frac{1}{F(p)} \right] \; + \;
\frac{1}{2} \, \frac{(\slsh{k}+\slsh{p})(k+p)^{\mu}}
{(k^2-p^2)}\left[ \frac{1}{F(k)}-\frac{1}{F(p)} \right] 
\nn\\&+&
   \frac{(k+p)^{\mu}}
{(k^2-p^2)}\left[ \frac{\M(k)}{F(k)}-
\frac{\M(p)}{F(p)}\right] \;. \label{Lvertex}
\eea
This choice is free of any unwanted kinematic singularities. What about 
the transverse part of the vertex? Does it also have some simple connection
with the fermion propagator or to expect and look for such a relation is
a wild goose chase? It is as important a question as its answer unclear.
Y. Takahashi, \cite{TWI1}, discovered generalized Ward identitues which put 
additional constraints on the transverse vertex~\cite{TWI2,TWI3,TWI4}. However, 
there is an alternative route based upon the arguments of multiplicative 
renormalizability in this connection. In a recent one loop perturbative calculation in arbitrary 
dimensions~\cite{BD2}, generalizing the results earlier known in 3 and 4 
dimensions,~\cite{BKP2,BKP1,CP1}, it has been shown that in the massless case 
in the limit when $k^2 >> p^2$, \footnote{Note that a typing mistake was done 
in Eq.~(26) of~\cite{BD2}. $\xi$ should not be there in the denominator.}
\bea
      \Gamma_T^{\mu}(k,p) = \frac{1}{2 k^2} \left[  
\frac{S_F^{-1}(k)}{\slsh{k}} - \frac{S_F^{-1}(p)}{\slsh{p}}  
\right] \; \left[ k^2 \gamma^{\mu} - k^{\mu} \slsh{k} \right] \;.
\eea
This is very much like the longitudinal vertex written in terms of the
fermion propagator, guided by the WGTI. There is of course no guarantee that
this relationship will survive higher order perturbative calculations.
However, if such relations exist, they will be of enormous help in
having a peep at the structure of the transverse vertex.

\section{Physical Observables}

We recall that the transverse part of the vertex plays a crucial rule to ensure the gauge 
independene of physical observables.  Therefore, it is appropriate at this
point to mention that without loss of generality, the transverse vertex 
can be expressed as~\footnote{We shall come back to the discussion of 
this basis in the light of perturbation theory and the kinematic
singularities.}: 
\be
\Gamma^{\mu}_{T}(k,p)=\sum_{i=1}^{8} \tau_{i}(k^2,p^2,q^2)T^{\mu}_{i}(k,p)
\;,\label{VT}
\ee
with an appropriate $\{T^\mu\}$ basis, defined first by Ball and 
Chiu,~\cite{BC}, as follows~:

\begin{eqnarray}
&T^{\mu}_{1}&=p^{\mu}(k\cdot q)-k^{\mu}(p\cdot q)\nonumber\\
&T^{\mu}_{2}&=\left[p^{\mu}(k\cdot q)-k^{\mu}(p\cdot q)\right]({\not\! k}
+{\not\! p})\nonumber\\
&T^{\mu}_{3}&=q^2\gamma^{\mu}-q^{\mu}{\not \! q}\nonumber\\
&T^{\mu}_{4}&=\left[p^{\mu}(k\cdot q)-k^{\mu}(p\cdot q)\right]k^{\lambda}
p^{\nu}{\sigma_{\lambda\nu}}\nonumber\\
&T^{\mu}_{5}&=q_{\nu}{\sigma^{\nu\mu}}\\
&T^{\mu}_{6}&=\gamma^{\mu}(p^2-k^2)+(p+k)^{\mu}{\not \! q}\nonumber\\
&T^{\mu}_{7}&=\frac{1}{2}(p^2-k^2)\left[\gamma^{\mu}({\not \! p}+{\not \! k})
-p^{\mu}-k^{\mu}\right]
+\left(k+p\right)^{\mu}k^{\lambda}p^{\nu}\sigma_{\lambda\nu}\nonumber\\
&T^{\mu}_{8}&=-\gamma^{\mu}k^{\nu}p^{\lambda}{\sigma_{\nu\lambda}}
+k^{\mu}{\not \! p}-p^{\mu}{\not \! k}\nonumber\\
\mbox{with}\;\;\;\;\;\;\;\;\;
&\sigma_{\mu\nu}&=\frac{1}{2}[\gamma_{\mu},\gamma_{\nu}]\qquad.
\end{eqnarray}

Gauge independence of physical observables should be the ultimate goal in non 
perturbative studies of SDE. If any proposal for the non perturbative form of the 
fermion-boson vertex is meant to be reliable, it would, certainly lead to 
gauge independent predictions for physical observables.

\subsection{Proposals for the Vertex}

Before going any further, it is important to present different proposals for the vertex 
found in literature both for QED4 and QED3.

\begin{itemize}

\item  {\bf {\underline{Curtis and Pennington:}}} 

The vertex proposed by Curtis and Pennington,~\cite{CP1}, in QED4 is
\bea
\nn
\Gamma^\mu_{CP} &=& \Gamma^\mu_{BC}+ \Gamma^{\mu T}_{CP} \;
\eea
where
\bea
\Gamma^{\mu T}_{CP} &=& \tau_6(k^2,p^2) T_6^{\mu}(k,p,q) \;
\eea
with
\bea
\tau_6(k^2,p^2) &=& 
\frac{1}{2}
\left[\frac{1}{F(k)}-\frac{1}{F(p)}\right] \; \frac{1}{d(k^2,p^2)}\;,
\eea
and
\bea
\nn
d(k^2,p^2)=\frac{(k^2-p^2)^2+[ {\cal M}^2(k)+ {\cal
M}^2(p)]^2}{k^2+p^2}\;.
\eea
In the massless case, it reduces to
\be
\nn
\tau_6= \frac{1}{2}\frac{k^2+p^2}{(k^2-p^2)^2}
\left[\frac{1}{F(k)} -\frac{1}{F(p)}\right]  \;.
\ee

\item {\bf {\underline{Bashir and Pennington I:}}} 

The following vertex was proposed by Bashir and Pennington~\cite{Bashirthesis}, 
also for QED4:
\bea
\nn
\Gamma^\mu_{BP_I} &=& \Gamma^\mu_{BC}+ \Gamma^{\mu T}_{BP_I} \;
\eea
where
\bea
\Gamma^{\mu T}_{BP_I} &=&  
\sum_{i=2,3,6,8} \tau_{i}(k^2,p^2)T^{\mu}_{i}(k,p)
\;
\eea
and, in the massless case, 
\bea
\nn
\tau_2(k^2,p^2) &=& a_2 \frac{1}{(k^2+p^2)(k^2-p^2)} \; 
\left[\frac{1}{F(k)} -\frac{1}{F(p)}\right]  \;,  \\
\nn
\tau_3(k^2,p^2) &=& a_3 \frac{1}{(k^2-p^2)} \; 
\left[\frac{1}{F(k)} -\frac{1}{F(p)}\right]  \;, \\
\nn
\tau_6(k^2,p^2) &=& a_6 \frac{(k^2+p^2)}{(k^2-p^2)^2} \; 
\left[\frac{1}{F(k)} -\frac{1}{F(p)}\right]  \;,  \\
\nn
\tau_8(k^2,p^2) &=& a_8 \frac{1}{(k^2-p^2)} \; 
\left[\frac{1}{F(k)} -\frac{1}{F(p)}\right]  \;,
\eea
with the constraints $1+a_2+2a_3+2a_8-2a_6=0$ and $a_3+a_6=1/2$, arising
by demanding multiplicative renormalizability of the massless fermion 
propagator. 
Choice
of Curtis and Pennington is $a_6=1/2, a_2=0$. The choice of Bashir and
Pennington is $a_6=-1/2$ and $a_2=2.75$. 

\item {\bf {\underline{Bashir and Pennington II:}}} 

A further generalization of the CP-vertex was carried out by
Bashir and Pennington,~\cite{BP1,BP2}, where, in the expansion
\bea
\Gamma^{\mu T}_{BP_{II}} &=&  
\sum_{i=2,3,6,8} \tau_{i}(k^2,p^2)T^{\mu}_{i}(k,p)
\;,
\eea
a more complicated structure for the $\tau_i(k^2,p^2)$ was proposed
in order to constrain the vertex to yield gauge independent answer
for $\alpha_c$ in QED4.

\item {\bf {\underline{Dong, Munczek and Roberts:}}}

Dong, Munczek and Roberts,~\cite{dong} made an attempt to propose
alternative choices in arbitrary dimensions which would also achieve
multiplicative renormalizability of the fermion propagator in
4 dimensions. 
\bea
\Gamma^{\mu T}_{DMR} &=&  
\sum_{i=2,3,6,8} \tau_{i}(k^2,p^2)T^{\mu}_{i}(k,p)  \;,
\eea
where
\bea
\tau_i &=& 
\frac{1}{k^2-p^2}\left[\frac{1}{F(k)}-\frac{1}{F(p)}\right]f^i\;.
\label{rel_tau_f}
\eea
The transverse coefficients $f^i$ are defined as~:
\bea
f^i&=&0 \;, \hspace{25mm}\mbox{for}\hspace{5mm}i\ne3,8  \nn\\
f^3&=&\frac{1}{2}\left(\frac{d}{2}-1\right)f^8  \;, \nn\\
f^8&=&\frac{1}{\frac{d}{2}-1}\frac{(d-1)I_3}{I_1-I_3}\;,
\eea
where $d$ is the number of space-time dimensions and
\bea
I_1&=&k^2p^2{\cal I}_1\nn\\
I_2&=&\frac{1}{2}\left((k^2+p^2) {\cal I}_1-1 \right)\nn\\
I_3&=&\frac{1}{2}[k^2+p^2]I_2\; ,
\eea
with ${\cal I}_n=\int d\Omega_d  {1}/{(k-p)^{2n}}$.

\item {\bf {\underline{Bashir, K{\i}z{\i}lers\"u and Pennington:}}}

 Bashir, K{\i}z{\i}lers\"u and Pennington~\cite{BKP1} present the most 
general non perturbative construction of the transverse vertex required by the
multiplicative renormalizability of the fermion propagator.

\item  {\bf {\underline{Burden and Roberts:}}} 

In the context of quenched QED3, Burden and Roberts have parametrized a slight 
modification to the BC-vertex in the following way,~\cite{BR2}~:
\bea
\Gamma^\mu_{BR}&=&\left[ a\frac{1}{F(k)}+(1-a)\frac{1}{F(p)}\right]
\gamma^\mu 
\nn\\ &&\nn\\
 &&
+ \frac{(k+p)^\mu((1-a)\slsh{k}-a\slsh{p})}{k^2-p^2}
\left[\frac{1}{F(k)}-\frac{1}{F(p)}\right]
\nn\\ &&\nn\\
&&
-\frac{(k+p)^\mu}{k^2-p^2}\left[ \frac{{\cal M}(k)}{F(k)}
-\frac{{\cal M}(p)}{F(p)}\right]\;,\label{BRvertex}
\eea
where $a$ is a free parameter.

\item  {\bf {\underline{Burden and Tjiang:}}} 

Using a similar reasoning, Burden and Tjiang,~\cite{BT1}, have deconstructed 
a one-parameter family of \emph{ans\"atze}  for massless QED3. 
\bea
\Gamma^{\mu T}_{BT} &=&  
\sum_{i=2,3,6,8} \tau_{i}(k^2,p^2)T^{\mu}_{i}(k,p)  \;,
\eea
Parametrizing the $\tau_i$ in exactly the same fashion as Dong, Munczek 
and Roberts,~\cite{dong}, and defining
\be
\bar{\tau}=\tau_8+(k^2+p^2)\tau_2\;,
\ee
and $\bar{f}$ identically, they choose
\bea
\bar{f}&=&-2(1+\beta)\frac{I(k,p)}{J(k,p)}\nn\\
f^3 &=&-\beta\frac{I(k,p)}{J(k,p)}\nn\\
f^6&=&0\;,
\eea
where
\bea
I(k,p)&=&\frac{(k^2+p^2)^2}{16kp}\ln{\left(\frac{(k+p)^2}{(k-p)^2}\right)}
-\frac{1}{4}(k^2+p^2)\nn\\ &&\nn\\
J(k,p)&=& \frac{(k^2-p^2)^2}{16kp}\ln{\left(\frac{(k+p)^2}{(k-p)^2}\right)}
-\frac{1}{4}(k^2+p^2)\;.
\eea
A choice of $\beta=1$ leads to the DMR-vertex. 

\item  {\bf {\underline{Bashir and Raya:}}}

More recently, an \emph{ansatz} for the fermion-boson vertex in massive QED3 
has been given by the authors~\cite{BashirR1}. Based on perturbation theory at one loop level, 
the transverse coefficients are found to have the form
\bea
  \hspace{-3mm} \tau_i(k,p) &=& \alpha g_i \left[ \sum_j^5
  a_{ij}(k,p) I(l_j)
+ \frac{a_{i6}(k,p)}{k^2 p^2}  \right]  \;, \hspace{5 mm} i=1, \cdots 8  \;,
\label{taui}
\eea
where $l_1^2 = \eta_1^2\chi/4$, $l_2^2 = \eta_2^2\chi/4$, $l_3^2 = k^2$, 
$l_4^2 = p^2$ and
$l_5^2 = q^2/4$. Functions $\eta_1$, $\eta_2$ and $\chi$ have been defined as
\bea
\chi&=&m_0^2(k^2-p^2)^2+q^2(m_0^2-k^2)(m_0^2-p^2) \;, \nn\\
\eta_1&=&- \left\{\frac{m_0^2(k^2-p^2)(2m_0^2-k^2-p^2)
+\chi}{2\chi(m_0^2-k^2)} \right\}   \;,      \nn\\
\eta_2&=&\left[ \frac
{\chi-m_0^2(k^2-p^2)(2m_0^2-k^2-p^2)}{\chi (m_0^2-p^2)}\right]\;.
\eea
The factors $g_i$ are $-g_1 = m_0 \Delta^2 g_2 = 2m_0 \Delta^2 g_3 = 
2 \Delta^2 g_4 = g_5 = 2 m_0 \Delta^2 g_6 = \Delta^2 g_7 = m_0 g_8 = m_0/ 4 
\Delta^2$, with $\Delta^2=(k \cdot p)^2 - k^2 p^2$. The coefficients 
$a_{ij}$ in the one loop perturbative expansion 
of the $\tau_i$, eq.~(\ref{taui}), are tabulated in~\cite{BashirR1}.
This allow us  to write a non perturbative form of $\tau_i$ as~:
\bea
\nn
&& \hspace{-8mm} \tau_i =
 g_i\Bigg\{ \sum_{j=1}^5 \left(
\frac{2a_{ij}(k,p)l_j^2}{\xi(m_0^2 + l_j^2)} \left[
\frac{\xi}{2  (\xi + 2) l_j^2  I(l_j)} \left(\frac{{\cal M}(l_j)}
{F(l_j)}
-m_0 \right)
- \left( 1-\frac{1}{F(l_j)} \right) \right] \right)
\\&&
 + \frac{2 a_{i6}(k,p)}{\xi\left[k^2A(p)-
p^2A(k)\right]}
\left[ \frac{1}{F(k)}-\frac{1}{F(p)} \right] \Bigg\} \;
\label{npvertexangle}
\eea
with the notation
\bea
I(p)&=&\frac{1}{\sqrt{-p^2}} \arctan{\sqrt{\frac{-p^{2}}{m_0^{2}}}}\;, \nn\\
A(p)&=& m_0- (m_0^2+p^2) I(p) \;.\label{Ip}
\eea
The transverse vertex is written solely as a function of the fermion 
propagator. Therefore, effectively, one has a WGTI for this part of the 
vertex. A computational difficulty to use the above vertex  in SDE studies 
could arise as the unknown functions $F$ and $\M$ depend on the angle between 
$k$ and $p$. This problem can be circumvented 
by defining an effective vertex which shifts the angular dependence
from the unknown functions $F$ and ${\cal M}$ to the known basic functions
of $k$ and $p$. This is done by re-writing 
eq.~(\ref{taui}), as follows~:
\begin{eqnarray}
     \tau_i(k,p) &=& \alpha g_i \left[ b_{i1}(k,p) I(k) + b_{i2}(k,p) I(p)
   + \frac{a_{i6}(k,p)}{k^2 p^2}  \right]  \;,
\end{eqnarray}
where
\begin{eqnarray}
      b_{i1}(k,p)  &=& a_{i1}(k,p) \, \frac{I(l_1)}{I(l_3)} + a_{i3}(k,p) 
+ \frac{1}{2} \, a_{i5}(k,p) \frac{I(l_5)}{I(l_3)} \;,    \\
 b_{i2}(k,p)  &=& a_{i2}(k,p) \, \frac{I(l_2)}{I(l_4)} + a_{i4}(k,p) 
+ \frac{1}{2} \, a_{i5}(k,p) \frac{I(l_5)}{I(l_4)} \;.
\end{eqnarray}
This form can be raised to a non-perturbative level exactly as before,
with the only difference that the functions $F$ and ${\cal M}$ are 
independent of the angle between the momenta $k$ and $p$~:
\begin{eqnarray}
\nonumber
&&\hspace{-8mm}
\tau_i = g_i\Bigg\{ \sum_{j=1}^2 \left( 
\frac{2b_{ij}(k,p)\kappa_j^2}{\xi(m_0^2 + \kappa_{j}^2)} \left[ 
\frac{\xi}{2 (\xi + 2) \kappa_j^2 I(\kappa_j)  } 
\left(\frac{{\cal M}(\kappa_j)}{F(\kappa_j)}-m_0 \right) 
\!-\!\left( 1-\frac{1}{F(\kappa_j)} \right) \right] \right)\\
&&  
+ \frac{2  a_{i6}(k,p)}{\xi\left[k^2 A(p)-
p^2A(k)\right]}
\left[ \frac{1}{F(k)}-\frac{1}{F(p)} \right] \Bigg\}, \nn \\
\end{eqnarray}
where $\kappa_1^2=k^2$ and $\kappa_2^2=p^2$. 
\end{itemize}

\subsection{Physical Observables in QED4}

In the previous section, we discussed several proposals for
the 3-point vertex in QED. Below we see how they perform to 
achieve gauge independent physical observables 
associated with the DCSB.

\begin{itemize}

\item  {\bf {\underline{The Bare Vertex:}}} 

       The use of the bare vertex in various gauges results in rather 
different values of the critical coupling above which fermion mass is 
generated, Figure~9. This is physically unacceptable.

\vspace*{-20pt}
\begin{center}
\epsfig{file=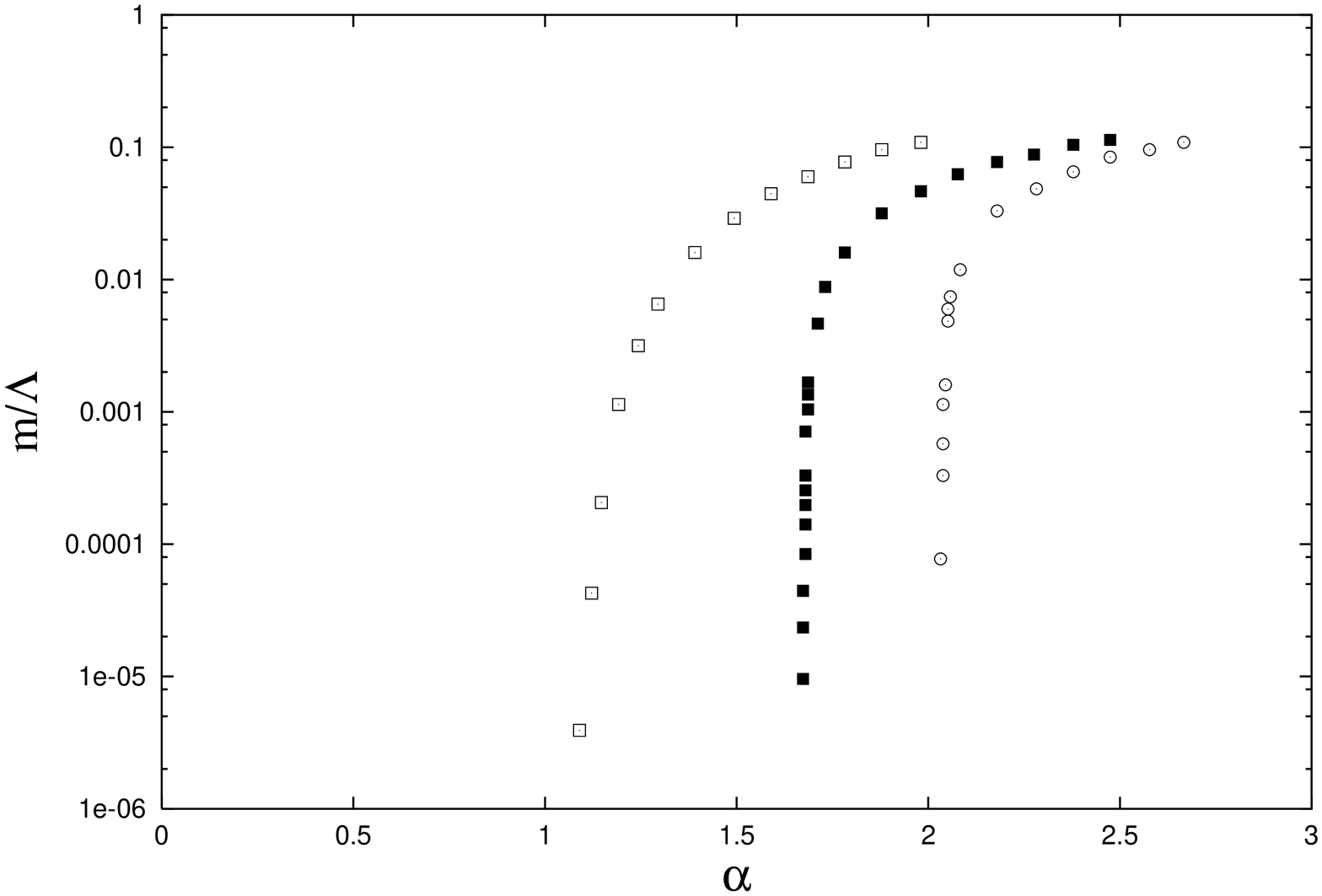,width=0.4\textwidth,angle=-90}
\end{center}
\begin{center}
{\sl Figure 9. Critical coupling in various gauges employing the
bare vertex.}
\end{center}

\item  {\bf {\underline{Curtis and Pennington:}}} 

Using the bifurcation analysis, one is led to the conclusion that in dramatic
contrast to the rainbow approximation, the critical coupling found
with the CP-vertex~\cite{CP1} is only weakly gauge dependent in the neighbourhood 
of the Landau gauge. The gauge dependence is reduced by about
$50\%$,~\cite{CP1}, Figure~10.

\begin{center}
\epsfig{file=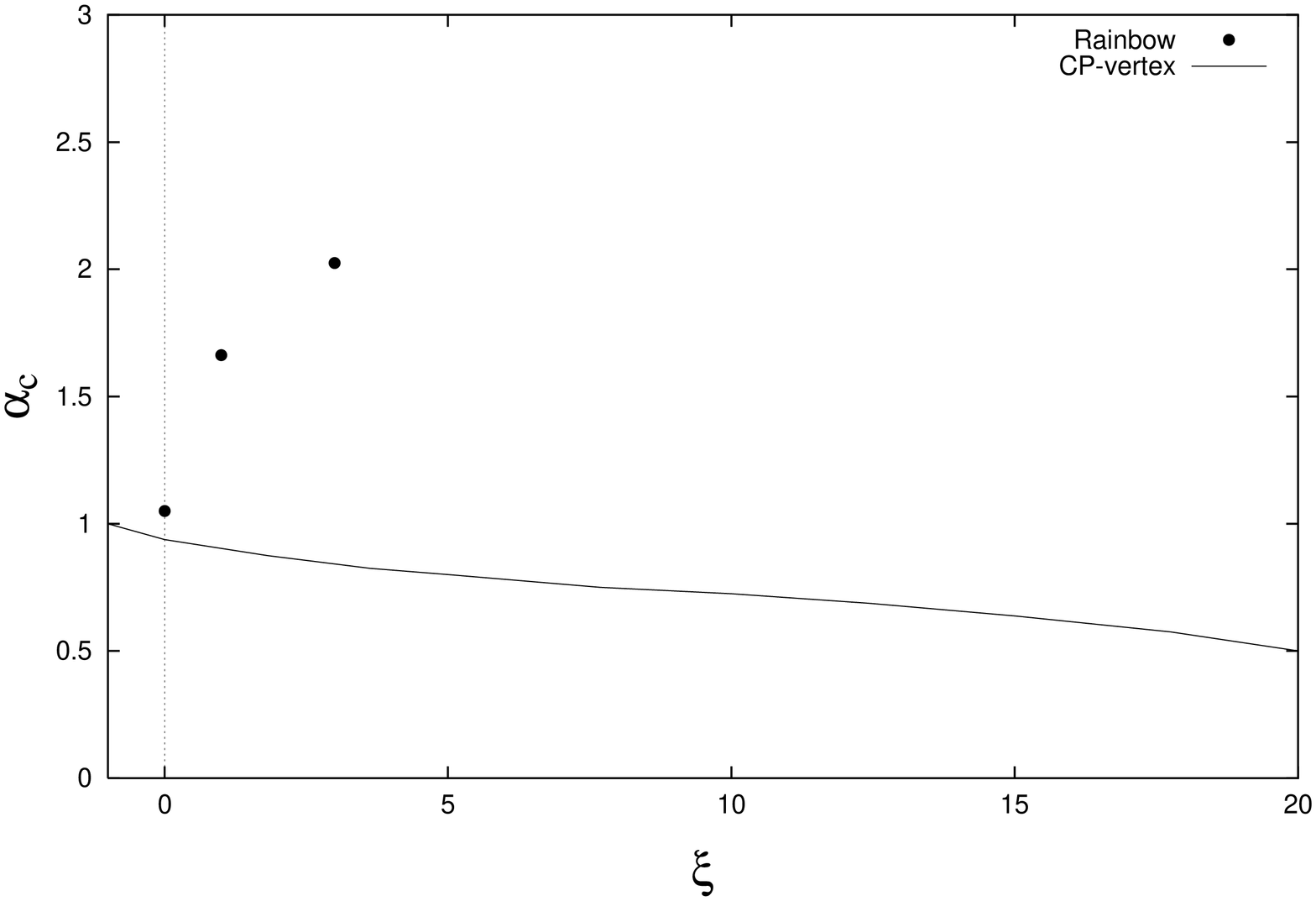,width=0.4\textwidth,angle=-90}
\end{center}
\begin{center}
{\sl Figure 10. Critical coupling in various gauges employing the
CP-vertex.}
\end{center}

\item  {\bf {\underline{Bashir and Pennington I:}}} 

In Figure~11, we plot the gauge dependence of $\alpha_c$ by making
use of the vertex proposed by Bashir and Pennington,~\cite{Bashirthesis}. 
For a comparison, the curve for the CP-vertex has also been plotted. 

\begin{center}
\epsfig{file=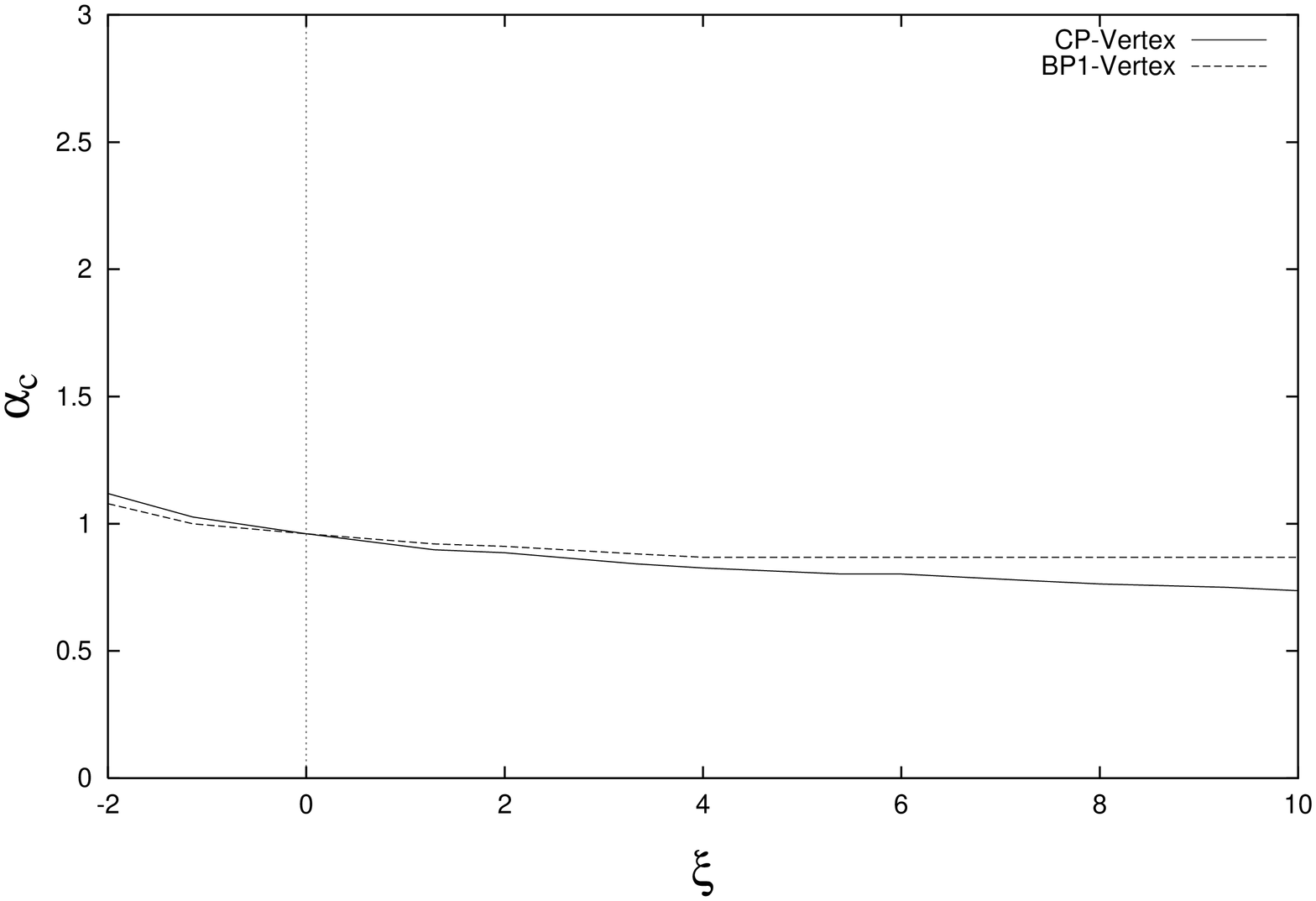,width=0.4\textwidth,angle=-90}
\end{center}
\begin{center}
{\sl Figure 11. Critical coupling in various gauges employing the
BP$_I$-vertex.}
\end{center}

One notes that the improvement not only exists in the neighbourhood of the
Landau gauge but that the curve continues to be flatter around 
$\alpha_c=0.93$ even up to quite large values of the gauge parameter.
Going from $\xi=0$ to $\xi=10$ reduces the gauge dependence by about 
$15\%$. The improvement becomes more significant when we are further away
from the Landau gauge,  Figure~12. For example, in going up to $\xi=70$, the 
change in $\alpha_c$ is improved by more than $60\%$ as compared to the
CP-vertex. These results are encouraging in the sense that one finds a vertex
which serves the aim of reducing the gauge dependence better than the
ones constructed before, in particular the CP-vertex, without introducing
any significant complication. But, however weak the variation of 
$\alpha_c$ with $\xi$ may be, any gauge dependence shows that the
 BP~I-vertex cannot be the exact choice. Even if it achieves a lot, it does not do it all.

\begin{center}
\epsfig{file=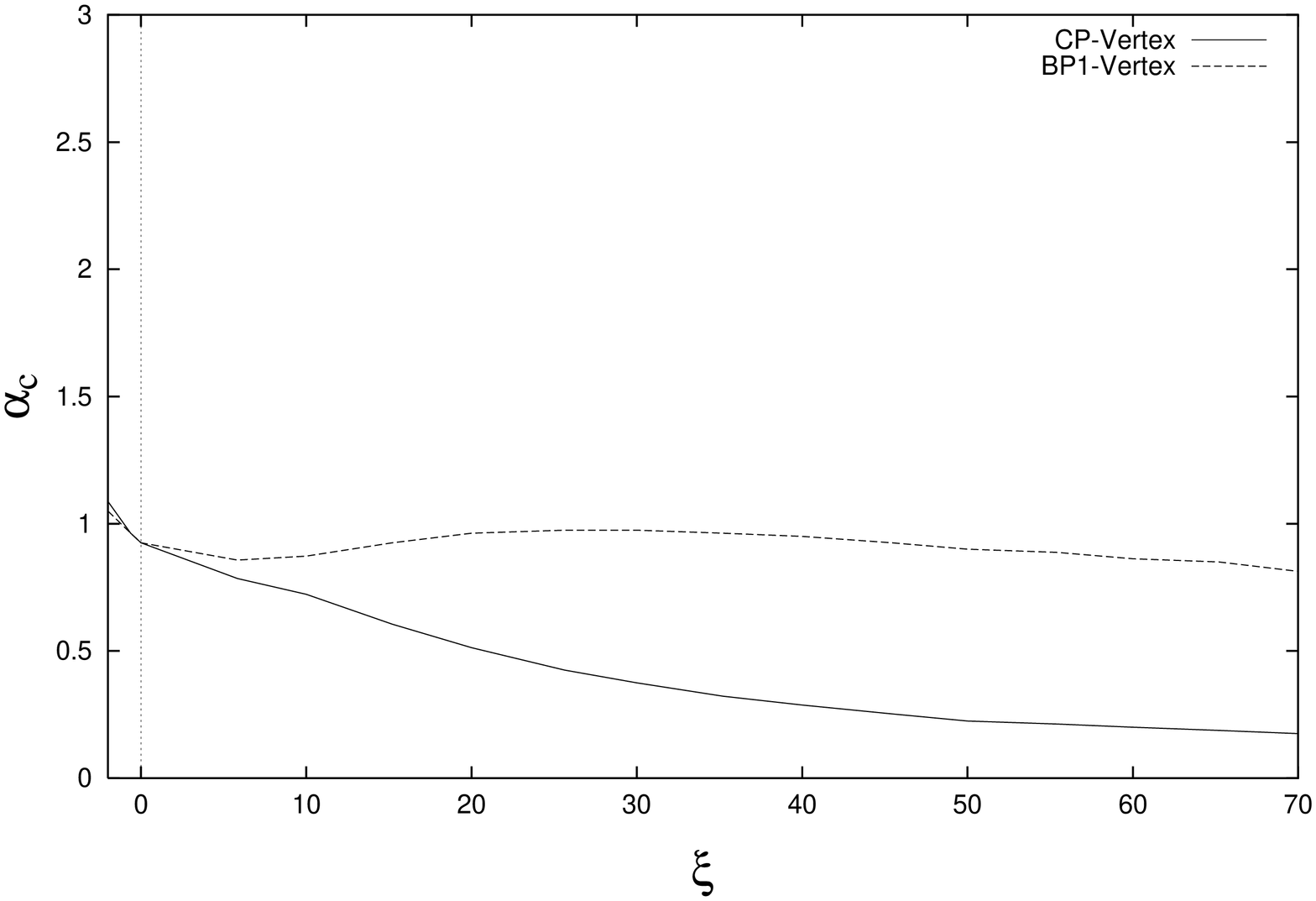,width=0.4\textwidth,angle=-90}
\end{center}
\begin{center}
{\sl Figure 12. Critical coupling in various gauges employing the
BP~I-vertex.}
\end{center}

\item {\bf {\underline{Bashir and Pennington II:}}}

The generalization of the vertex~\cite{Bashirthesis} developed in~\cite{BP1,BP2}
{\em renders the chiral symmetry breaking phase transition gauge independent in 
addition to ensuring that the massless fermion propagator is gauge covariant and multiplicatively
renormalizable and also that the WGTI is satisfied.} 

\end{itemize}

\subsection{Physical Observables in QED3}

In QED3, the coupling has dimensions of mass. As a result
there is no critical 
value of coupling for the DCSB to take place. If it happens at one value of
coupling, it takes place at every other value as well. The physical 
observables related to the DCSB are the the chiral condensate and 
the physical fermion mass. We review these quantities as predicted by the use of different vertex ans\"atze.

\begin{itemize}

\item  {\bf {\underline{The Bare Vertex:}}} 

The gauge dependence of 
the chiral condensate using the bare vertex was studied by 
Burden and Roberts~\cite{BR2} in a narrow vicinity of the Landau gauge,
namely from $\xi=0$ to $\xi=1$, where it seems to depend considerably on the
choice of the gauge parameter. Later Bashir, Huet and Raya,~\cite{BHR},
 studied this
dependence in a wider range of values of the gauge parameter, i.e.,
from $\xi=0$ to $\xi=5$. They find that with the increasing value of the
gauge parameter, the gauge dependence gets reduced.

\item  {\bf {\underline{Burden and Roberts:}}}

A slight modification to the Ball-Chiu vertex proposed by Burden and 
Roberts,~\cite{BR2},  eq.~(\ref{BRvertex}), allows them to tune the parameter 
$a$ to achieve a nearly gauge independent chiral condensate in the region
$\xi=0$ to $\xi=1$. The value they report is $a=0.53$ for this purpose.

\item  {\bf {\underline{Curtis and Pennington:}}}

Strictly speaking,  Curtis-Pennington vertex was discovered only for the
4-dimensional case. However, if one uses it to calculate the condensate in
quenched QED3 in various gauges,~\cite{FRDM}, one gets the upper curve shown 
in Fig.~13.
\end{itemize}

\begin{center}
\epsfig{file=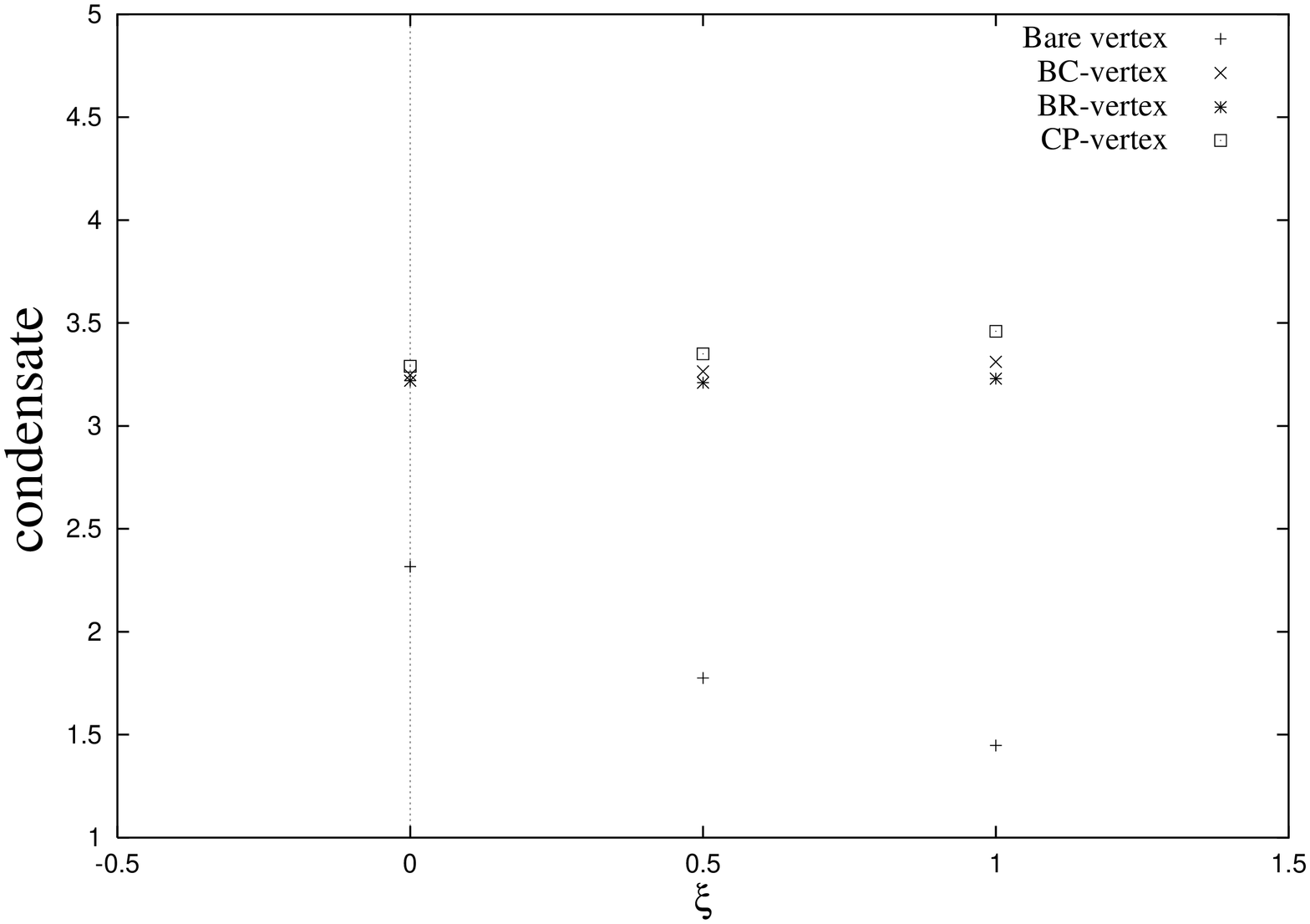,width=0.4\textwidth,angle=-90}
\end{center}
\begin{center}
{\sl Figure 13. The condensate in quenched QED3 with various vertex
{\em ans\"atze}.}
\end{center}

\section{SDEs and Dimensional Regularization}

In the hunt for the non perturbative form of the fermion-boson vertex, the root of gauge invariance violation can be found in several places. We have earlier commented on the restrictions imposed on the vertex, but so far we have left an issue apart.  In QED4, it is a common practice to use an ultraviolet cut-off to regulate the divergent integrals. 
Such procedure, when employed carelessly, violates translational invariance
(in momentum space) which leads to a violation of gauge invariance in the 
final results. This can be exemplified by calculating the one loop correction to the massless 
fermion propagator, which explicitly reads
\bea
\frac{1}{F(p)}&=&1+\frac{i\alpha}{4\pi^3 p^2} \int_M \frac{d^4k}{k^2q^4} 
\Bigg\{ -3k\cdot p(k^2+p^2)+4(k\cdot p)^2 +2k^2p^2 
\nn\\&&
+ \xi(k^2 p\cdot q -p^2 k\cdot q)\Bigg\}\;.
\eea
Using an ultraviolet cut-off to regularize the integral, one finds
\be
\frac{1}{F(p)}= 1-\frac{\alpha\xi}{4\pi}\ln\frac{p^2}{\Lambda^2}+
\frac{\alpha\xi}{8 \pi}\;.
\ee
The last term of this expression is spurious. It is known that such term can 
be removed by first making use of the WGTI in the SDE in the $\xi$-term and 
then carrying out the angular integration~\cite{CP3}. 
However, the fact remains that this method is ad hoc and it would be 
essential
to confirm its validity.
A reliable scheme of regularization of the divergent integrals is the 
dimensional regularization scheme.
This approach is manifestly gauge invariant. These doubts were raised and
addressed in~\cite{SSW,GSSW,Ay}.
However, the results obtained within dimensional regularization agree with 
those obtained by employing a correctly implemented cut-off within the 
numerical precision, and
the numerical difficulties encountered in implementing this regulator
has limited its further use, at least for the time being.
It may be worth reminding at this point that QED3 lacks
 ultraviolet divergences. Therefore, it serves as an excellent laboratory
to study the questions of gauge invariance without worrying about
the procedure of regularization~\cite{BHR}.

\section{Kinematic Singularities}

As mentioned before, the full vertex should be independent of unwanted
kinematic singularities which are absent in perturbation theory.
The WGTI is non singular as $k\to p$, and hence, the construction of the 
longitudinal piece of the vertex as proposed by Ball and Chiu ensures 
this condition is fulfilled by this part of the vertex. 
As a result, it also has no singularities when $k^2 \rightarrow p^2$.
In the context of the longitudinal construction of vertices, the vertex
proposed by  Atkinson, Johnson, and Stam~\cite{AJS} exhibits kinematic 
singularities, whereas  the one of Haeri~\cite{Haeri} is free of them.

The transverse part of the vertex should also be independent of such 
singularities. It is natural to expect that a good choice for the
basis to express the  transverse part of the vertex is one that 
ensures that each coefficient is free of kinematic singularities.  
The original basis proposed by Ball and Chiu~\cite{BC} was designed
so that this objective is achieved  at the one loop order in the Feynman gauge $(\xi=1)$. 
Later a similar perturbative calculation in an arbitrary covariant gauge
by  K{\i}z{\i}lers\"u, Reenders and Pennington~\cite{Reenders} revealed
that the coefficients of this basis develop kinematic singularities.
They proposed a straightforward modification of the earlier basis that 
ensures each transverse component is separately free of
kinematic singularities in any covariant gauge. The only modification
takes place in the definition of $T_4^{\mu}$ which now reads~:
\begin{eqnarray}
T^{\mu}_4&=&q^2\left[\gamma^{\mu}(\not\!p+\not\!k)-p^{\mu}-k^{\mu}\right]
+2(p-k)^{\mu}k^{\lambda}p^{\nu}\sigma_{\lambda\nu} \;.
\end{eqnarray}
This is also true in QED3, as it was demonstrated in Ref.~\cite{BashirR1}, 
and in fact also in arbitrary dimensions which can be checked after a proper
conversion of basis in the work of Davydychev 
\emph{et. al.}~\cite{davydychev}.

\begin{itemize}

\item  {\bf {\underline{Curtis and Pennington:}}}

This widely implemented vertex \emph{ansatz} does not exhibit kinematic 
singularities for massive fermions for which the vertex
was proposed.

\item  {\bf {\underline{Burden and Roberts:}}}

As a slight modification to the Ball-Chiu vertex, this \emph{ansatz} is free of 
kinematic singularities.

\item  {\bf {\underline{Dong, Munczek, and Roberts:}}}

This \emph{ansatz} exhibits logarithmic singularities in 3 dimensions.

\item  {\bf {\underline{Burden and Tjiang:}}}

This deconstruction of a family of vertex \emph{ans\"atze}
possesses logarithmic singularities.

\item  {\bf {\underline{Bashir and Pennington I \& II:}}}

This vertex proposal and its generalization do not exhibit kinematic 
singularities.

\item  {\bf {\underline{Bashir and Raya:}}}

This family of vertex \emph{ans\"atze} was constructed free of kinematical singularities.
\end{itemize}

\section{Transformation Under C, P and T}

Another constraint on the non perturbative form of the fermion-boson 
vertex arises from the requirement that it should transform in the same way the bare 
vertex does under C, P, and T.  Thus, e.g., under the charge conjugation 
operation, we must have~:
\bea
     C \Gamma_{\mu} C^{-1} &=& - \Gamma_{\mu}^{{\rm T}} (-p,-k) \;, \nn
\eea
Now making use of the 
identities
\bea
   C=-C^{\rm T}  \hspace{20mm} C \gamma_{\mu} C^{-1} = 
- \gamma_{\mu}^{{\rm T}} \;, \nn
\eea
it is easy to see that for the Ball and Chiu selection of the transverse
basis vectors,
\bea
       C T_i^{\mu}(k,p)C^{-1} &=& - T_i^{\mu {\rm T}}(-p,-k)   
\hspace{15mm} {\rm for} \hspace{5mm} i \neq 6   \nn \\
 C T_6^{\mu}(k,p)C^{-1} &=&  T_6^{\mu {\rm T}}(-p,-k) \;.
\eea
This implies that all the $\tau_i(k^2,p^2)$ are symmetric, except for 
$\tau_6(k^2,p^2)$ which is anti-symmetric under $k^2 \leftrightarrow p^2$.
With the choice of the transverse basis as proposed by K{\i}z{\i}lers\"u, 
Reenders and Pennington~\cite{Reenders}, correct charge conjugation 
properties demands all of the $\tau_i$ to be symmetric under 
$k\leftrightarrow p$, except for $\tau_4$ and $\tau_6$ which should be 
antisymmetric. Most of the vertices proposed so far fulfil this
requirement by construction,~\cite{AJS,BC,Haeri,CP1,BP1,BP2,dong,BT1},
except the one presented by Burden and 
Roberts,~\cite{BR2},~eq.~(\ref{BRvertex}). In this equation if we substitute
$a={1}/{2}+\delta$, we observe that the BR-vertex can 
also be written as~:
\bea
\Gamma^\mu_{BR}&=&\Gamma^\mu_{BC}+\delta\left[\frac{1}{F(k)}-\frac{1}{F(p)}
\right][k^2-p^2]\left\{\gamma^\mu(p^2-k^2)+(k+p)^\mu\slsh{q}\right\}\nn\\
&=&\Gamma^\mu_{BC}+\delta\tau_6 T^\mu_6\;,
\eea
with
\be
\tau_6=(k^2-p^2)\left[\frac{1}{F(k)}-\frac{1}{F(p)}\right]\;.
\ee
As already mentioned, charge conjugation symmetry for the vertex requires 
$\tau_6$ to be 
antisymmetric under the interchange of $k$ and $p$, but the BR-vertex forces
it to be symmetric.

\section{Perturbation Theory}

Interactions determine the dynamical structure of any theory. In the case of 
QED, interactions correspond to the fermion-boson vertex. The SDE for
this vertex and the higher-order Green functions are very complicated
and a non perturbative  solution of these equations is a formidable task.
As we have explicitly seen in previous sections,
it is customary to make an \emph{ansatz} for the vertex to solve the 
SDE for the fermion propagator. Therefore, an intelligent guess
of its non perturbative form is a key issue. On the other hand,
we know that perturbation theory (PT) is the only fully reliable scheme of 
truncation of the SDEs where all the key features of a gauge field theory, 
such as the gauge identities, are satisfied automatically at every order of
approximation. Therefore, to stand a chance of maintaining these 
features at the non perturbative level, a connection with PT could play a vital
role. Besides, for physically meaningful solutions of these equations, 
we must demand the results from SDEs to agree with perturbative ones in the 
weak coupling regime. Therefore, PT can guide us in the search for the 
possible non perturbative forms of the fermion-boson interaction.

For the study of DCSB, the captivatingly simple use of the rainbow 
approximation is based on the lowest order perturbative expansion of the
vertex. However, not surprisingly, it  violates gauge invariance and its 
implications at the non perturbative level. Thus, for reliable results, we
must look for more 
sophisticated \emph{anz\"atze} which do respect these key features, peeping 
at the higher orders in perturbation theory. Recall that the full 
fermion-boson vertex can be expressed in terms of the twelve amplitudes 
given in eq.~(\ref{FVdef}), which would suggest that it 
involves the same number of independent functions. However the non 
perturbative form of these functions should be related to the 
fermion propagator in the way dictated by the WGTI.  Of course this identity 
constrains only the so called longitudinal part of the vertex, and hence 
only some of the twelve functions which define the full three-point function. Additional 
constraints can be set by demanding these functions to fulfil their 
corresponding LKFT, making sure that the dynamically generated fermion 
propagator does so as well. As in PT this is achieved automatically,
one cannot resist the temptation to be guided by it into the non perturbative
regime. The procedure  is as follows~: the longitudinal part of the vertex can be 
constrained by the perturbative expansion of the fermion propagator by means 
of the WGTI in terms of the Ball-Chiu vertex \emph{ansatz}, 
eq.~(\ref{Lvertex}). On the other hand, the perturbative calculation of the fermion-boson 
vertex can allow us to identify the transverse part of the vertex by subtracting from 
it the previously found longitudinal piece. As pointed out in~\cite{Reenders}, 
this procedure should be performed in an arbitrary gauge, otherwise we will be 
clueless of the non perturbative form of the vertex. For instance, if we 
calculated the vertex in massless QED 
in the Landau gauge, we would find that the $v_1(k,p)$ function was like its 
bare form, just $\gamma^\mu$, and we would have no information whatsoever of its 
non perturbative $\frac{1}{2}[1/F(k)+1/F(p)]\gamma^\mu$ structure.
In the spirit of this reasoning, the analysis at one loop level was already
performed in literature. Based on PT, we can compute the fermion propagator 
at the one loop level from the diagram
\begin{center}
\vspace*{-1.5cm}
\begin{picture}(500,100)(0,0)
\SetScale{0.7}
\ArrowLine(50,50)(150,50)
\CCirc(100,50){3}{}{}
\PText(145,60)(0)[]{-1}
\PText(100,45)(0)[]{p}
\ArrowLine(200,50)(300,50)
\PText(295,60)(0)[]{-1}
\PText(250,45)(0)[]{p}
\Vertex(430,50){1} \Vertex(370,50){1}
\ArrowLine(350,50)(450,50)
\PhotonArc(400,50)(30,0,180){4}{8.5}
\LongArrowArc(400,50)(20,60,120)
\PText(400,45)(0)[]{k}
\PText(400,95)(0)[]{q}
\PText(175,52)(0)[]{=}
\PText(325,52)(0)[]{-}
\end{picture}\\
\vspace{-10pt}
{\sl Figure~14~: One loop correction to the fermion propagator.}
\end{center}
which yields the following expressions for $F(p)$ and $\M(p)$
\bea
\frac{1}{F(p)}&=&\frac{e^2}{i(2\pi)^{d}}\frac{(d-2)\xi}{2p^2}
\left[(p^2+m^2)\int \frac{d^dw}{(k-w)^2(w^2-m^2)}-\int \frac{d^dw}{w^2-m^2}
\right],\nn\\
\frac{{\cal M}(p)}{F(p)}&=&\frac{e^2}{i(2\pi)^{d}}m(d-1-\xi)
\int \frac{d^dw}{(k-w)^2(w^2-m^2)}\;.
\eea
 Then, from the expression~(\ref{Lvertex}), we can set perturbative 
 constraints to the longitudinal part of the vertex, thus achieving the first 
 step of the analysis. 
The second step demands calculating the one loop 
correction to the fermion-boson vertex, which can be done from the diagram
\begin{center}
\vspace{-1.5cm}
\SetScale{0.7}
\begin{picture}(500,100)(0,0)
\ArrowLine(125,75)(75,50)
\ArrowLine(75,50)(125,25)
\Vertex(75,50){1}
\Photon(25,50)(75,50){-3}{4}
\CCirc(75,50){3}{}{}
\PText(50,63)(0)[]{q}
\LongArrow(55,65)(45,65)
\PText(100,72)(0)[]{k}
\PText(100,35)(0)[]{p}
\ArrowLine(275,75)(225,50)
\ArrowLine(225,50)(275,25)
\Vertex(225,50){1}
\Photon(175,50)(225,50){-3}{4}
\PText(200,63)(0)[]{q}
\LongArrow(205,65)(195,65)
\PText(250,72)(0)[]{k}
\PText(250,35)(0)[]{p}
\ArrowLine(425,75)(375,50)
\ArrowLine(375,50)(425,25)
\Vertex(375,50){1}
\Photon(375,50)(325,50){-3}{4}
\Photon(410,68)(410,32){3}{4}
\PText(350,63)(0)[]{q}
\LongArrow(355,65)(345,65)
\PText(395,72)(0)[]{k-w}
\PText(395,35)(0)[]{p-w}
\PText(418,50)(0)[]{w}
\LongArrow(425,45)(425,55)
\PText(150,52)(0)[]{=}
\PText(300,52)(0)[]{-}
\end{picture}\\
\vspace{-10pt}
{\sl Figure~15~: One loop correction to the vertex.}
\end{center}
and can be expressed as~:
\be
\Gamma^{\mu}(k,p)=\,\gamma^{\mu}+\,\Lambda^{\mu} \;.  \label{01loopvertex}
\ee
The one loop correction can be written as
\bea
\Lambda^\mu &=& -4 i\pi \alpha \int \frac{d^dw}{(2\pi)^d} \left\{ \frac{A^\mu}{[(p-w)^2-m^2][(k-w)^2-m^2]w^2} 
\right. \nn\\&& +  \left. 
\frac{B^\mu}{[(p-w)^2-m^2][(k-w)^2-m^2]w^4} \right\}\;,
\eea
where
\bea
A^\mu &=& \gamma^\alpha (\slsh{p}-\slsh{w}) \gamma^\mu 
(\slsh{p}-\slsh{w}) \gamma_\alpha 
\nn\\&&
 + m \gamma^\alpha[(\slsh{p}-\slsh{w})\gamma^\mu+\gamma^\mu 
(\slsh{k}-\slsh{w})]\gamma_\alpha+m^2 \gamma^\alpha \gamma^\mu 
\gamma_\alpha\nn\\
B^\mu &=& \slsh{w}(\slsh{p}-\slsh{w})\gamma^\mu (\slsh{k}-\slsh{w}) 
\slsh{w} 
\nn\\&&
+ m \slsh{w}[(\slsh{p}-\slsh{w})\gamma^\mu+\gamma^\mu 
(\slsh{k}-\slsh{w})] \slsh{w}+m^2 \slsh{w}\gamma^\mu \slsh{w}\;.
\eea
The master integrals to be calculated in this connection are~:
\bea
 K^{(0)}&=&\int_{M}\,d^dw\,\frac{1}{[(p-w)^2-m^2]\,[(k-w)^2-m^2]}
\nn \\
J^{(0)}&=&\int_{M}\,d^dw\,\frac{1}{w^2\,[(p-w)^2-m^2]\,[(k-w)^2-m^2]}
\nn \\
J^{(1)}_{\mu}&=&\int_{M}\,d^dw\,
\frac{w_{\mu}}{w^2\,[(p-w)^2-m^2]\,[(k-w)^2-m^2]}
\nn \\
J^{(2)}_{\mu\nu}&=&\int_{M}\,d^dw\,
\frac{w_{\mu}w_{\nu}}{w^2\,[(p-w)^2-m^2]\,[(k-w)^2-m^2]}
\nn \\
I^{(0)}&=&\int_{M}\,d^dw\,
\frac{1}{w^4\,[(p-w)^2-m^2]\,[(k-w)^2-m^2]}
\nn \\
I^{(1)}_{\mu}&=&\int_{M}\,d^dw\,
\frac{w_{\mu}}{w^4\,[(p-w)^2-m^2]\,[(k-w)^2-m^2]}
\nn \\
 I^{(2)}_{\mu\nu}&=&\int_{M}\,d^dw\,
\frac{w_{\mu}w_{\nu}}{w^4\,[(p-w)^2-m^2]\,[(k-w)^2-m^2]}
\label{integrals}  \;.
\eea
There exist several works which carry out this calculation employing
a varying degree of complication~:

\begin{itemize}

\item Ball and Chiu,~\cite{BC}, evaluate the transverse vertex at the one 
loop level in the Feynman gauge for massive fermions (QED4).

\item Curtis and Pennington,~\cite{CP1},
evaluate the transverse vertex at the one loop level
in an arbitrary covariant gauge in the limit when momentum in one of the
fermion legs is much greater than in the other leg for massless fermions 
(QED4). 

\item K{\i}z{\i}lers\"u, Pennington and Reenders,~\cite{Reenders}, 
evaluate the 
transverse vertex at 
the one loop level in an arbitrary covariant gauge for all regions of momenta
for massive fermions (QED4).

\item Bashir, Pennington and K{\i}z{\i}lers\"u,~\cite{BKP1}, evaluate the 
transverse 
vertex at the one loop level in an arbitrary covariant gauge for all regions 
of momenta for massless fermions (QED3).

\item Bashir and Raya,~\cite{BashirR1}, evaluate the transverse 
vertex at the one loop level in an arbitrary covariant gauge for all regions 
of momenta for massive fermions (QED3).

\item Davydychev, Osland and Sakset,~\cite{davydychev}, evaluate the 
transverse vertex at the one loop level in an arbitrary covariant gauge for 
all regions of momenta in massive QED in arbitrary dimensions.

\end{itemize}

\begin{itemize}

\item  {\bf {\underline{Case $d=4$}}} 

In the four dimensional case, the one loop perturbative 
correction to the fermion propagator was calculated in~\cite{Reenders}. 
The authors find
\bea
F(p)&=&1-\frac{\alpha\xi}{4\pi}
\left[ C\mu^\epsilon+\left(1-\frac{m^2}{p^2}\right) (1-L)\right] \;,
 \label{KRPF} \\
{\cal M}(p)&=&m+\frac{\alpha m}{\pi}
\left[\left(1+\frac{\xi}{4}\right)+\frac{3}{4}(C\mu^\epsilon-L)
+\frac{\xi}{4}\frac{m^2}{p^2}(1-L)\right] \label{KRPMC} \;,
\eea
where
\bea
L&=&\left(1+\frac{m^2}{p^2} \right) \ln{\left(1+\frac{p^2}{m^2}\right)}
\;,
\nonumber \\ 
C&=&-\frac{2}{\epsilon}-\gamma-\ln{\pi}-\ln{\left(\frac{m^2}{\mu^2}\right)}
\;.\label{KRPfunctions}
\eea
These results, in combination with eq.~(\ref{Lvertex}) yield the following 
perturbative expression for the one loop longitudinal vertex
\bea
\Gamma_L^\mu(k,p)&=&\frac{\alpha\xi}{8 \pi} \gamma^\mu 
\left[ 2 C \mu^\epsilon +\left( 1+\frac{m^2}{k^2}\right) (1-L')
+\left( 1+\frac{m^2}{p^2}\right) (1-L)\right]\nn\\
&&
\hspace{-2cm}
+\frac{\alpha\xi}{8 \pi}  \frac{(k+p)^\mu(\slsh{k}+\slsh{p})}{(k^2-p^2)} 
 \left[ m^2\left( \frac{1}{k^2} - \frac{1}{p^2}\right) -
\left( 1+\frac{m^2}{k^2}\right) L'+
\left( 1+\frac{m^2}{p^2}\right)L\right]
\nn\\
&&
\hspace{-2cm}
+\frac{\alpha m}{4 \pi} (3+\xi) 
\frac{(p+k)^\mu}{(k^2-p^2)}[L-L']\;,\label{longitudinal4d}
\eea
where $L' = L(k\leftrightarrow p)$.
On the other hand, the one loop correction to the vertex is found to be 
written as combination of the functions defined in eq.~(\ref{KRPfunctions}), 
of the function
\be
S = \frac{1}{2}\sqrt{1-4\frac{m^2}{q^2}} \ln{\left| 
\frac{\sqrt{1-4m^2/q^2}+1}{\sqrt{1-4m^2/q^2}-1}\right|}\;,
\ee
and $J_0$ which is a combination of Spence functions. We refer the reader 
to the original reference, \cite{Reenders} for the explicit lengthy expressions.
A subtraction of eq.~(\ref{longitudinal4d}) from this correction gives the following 
expression for the transverse vertex
\bea
\Gamma^\mu_T(k,p)&=& \frac{\alpha}{4\pi} \Bigg\{ \sum_{i=1}^{12} V_i^\mu 
\left(\frac{1}{2\Delta^2} [a_1^{(i)}-(\xi-1)a_2^{(i)}] J_A\right.
\nn\\&&
+ \frac{1}{2\Delta^2} [b_1^{(i)}-(\xi-1)b_2^{(i)}] J_B
\nn\\&&
+\frac{1}{2\Delta^2} [c_1^{(i)}-(\xi-1)c_2^{(i)}] I_A\nn\\
&& + \frac{1}{2\Delta^2} [d_1^{(i)}-(\xi-1)d_2^{(i)}] I_B 
\nn\\&&
+\frac{1}{2p^2(k^2-p^2)(p^2-m^2)\Delta^2} [e_1^{(i)}-(\xi-1)e_2^{(i)}] 
L
\nn\\&&
+\frac{1}{2k^2(k^2-p^2)(k^2-m^2)\Delta^2} [f_1^{(i)}-(\xi-1)f_2^{(i)}] 
L'
\nn\\&&
 + \frac{1}{\Delta^2} [g_1^{(i)}-(\xi-1)g_2^{(i)}] S \nn\\
&& + \frac{1}{2\Delta^2} [h_1^{(i)}-(\xi-1)h_2^{(i)}] J_0 
\nn\\&&
 +\left. \frac{1}{\Delta^2} [l_1^{(i)}-(\xi-1)l_2^{(i)}]\right) \Bigg\} \;,
\eea
where $\Delta=(k\cdot p)^2-k^2p^2$. The expressions $J_A, J_B, I_A$ and $I_B$ 
are combinations of $L,L',S$ and $J_0$, and along with the coefficients 
$a_i,b_i,c_i,d_i,e_i,f_i,g_i,h_i$ and $l_i$ are tabulated in 
Ref.~\cite{Reenders}. It 
yields the results obtained by~\cite{BC} and~\cite{CP1} as special cases.

\item  {\bf {\underline{Case $d=3$}}} 

For QED3, the same exercise was done in~\cite{BashirR1} for the
massive fermions. The authors find 
\bea
\frac{1}{F(p)}&=&1-\frac{\alpha\xi}{2 p^2}  \,
\left[ m - (m^2+p^2) \, I(p)
\right] \;,\nn\\
\frac{\M(p)}{F(p)}&=&m\left[1+
\alpha (\xi+2) \, I(p) \right] \;, \label{FMradial}
\eea
with $I(p)$ defined as in eq.~(\ref{Ip}). With the aid of these expressions, 
the one loop  longitudinal part of the 
vertex can be written as
\bea
   \Gamma^{\mu}_{L} &=& \left[ 1 + \frac{\alpha \xi}{4} \, \sigma_1 \right]
    \, \gamma^{\mu} \; + \;  \frac{\alpha \xi}{4} \, \sigma_2 \, \left[
{k^{\mu}}\slsh{k} \, +  \, {p^{\mu}}\slsh{p} \, + \,
{k^{\mu}}\slsh{p} \, +  \, {p^{\mu}}\slsh{k}  \right]
\nn \\ &+&
\alpha (\xi+2) \sigma_3 \, \left[ k^{\mu} + p^{\mu}  \right]\;,
\label{1loopLvertex}
\eea
where
\bea
\sigma_1 &=&   \frac{m^2+k^2}{k^2}  \, I(k) \; + \;
 \frac{m^2+p^2}{p^2}  \, I(p) \; - \; m \frac{k^2+p^2}{k^2 p^2} \;,
  \nn \\
\sigma_2 &=& \frac{1}{k^2-p^2} \, \left[
 \frac{m^2+k^2}{k^2}  \, I(k) \; - \;
 \frac{m^2+p^2}{p^2}  \, I(p) \; + \; m \frac{k^2-p^2}{k^2 p^2}
  \right] \;,  \nn \\
\sigma_3 &=& m \;  \left[ I(k) \; - \; I(p)  \right]   \;.
\label{sigmas}
\eea
 The one loop correction for the vertex in this case is more simple, since it 
can be written in terms of functions $I(p)$, with 
complicated arguments though. For the identification of the transverse part 
of the vertex, the authors made use of the decomposition (\ref{VT}) for the
vertex, making use of the basis vectors proposed by K{\i}z{\i}lers\"u \emph{et. 
al.}
After a lengthy but straightforward algebra, the coefficients 
$\tau_{\it i}$ are identified, eq.~(\ref{taui}). The main advantage
of QED3 is that no special functions are involved and raising the perturbative
expression to a no perturbative status is possible.

\item  {\bf {\underline{Case of arbitrary dimensions}}} 

  Davydychev \emph{et. al.}, \cite{davydychev}, have calculated the one loop 
correction to the fermion propagator and to the fermion-boson vertex. These 
authors introduce two  Master Integrals~:
\bea
I(\nu_1,\nu_2,\nu_3)&\equiv&\int\frac{d^dw}
{[(p-w)^2-m^2]^{\nu_1}[(k-w)^2-m^2]^{\nu_2}[w^2]^{\nu_3}},\\
{\cal I}(\nu_1,\nu_2,\nu_3)&\equiv&\int\frac{d^dw}
{[(p-w)^2]^{\nu_1}[(k-w)^2]^{\nu_2}[w^2-m^2]^{\nu_3}}\;.
\eea
In terms of these integrals, one can write out the fermion propagator and 
fermion-boson vertex in the context of QED in arbitrary dimensions.
They find that the fermion propagator at one loop order can be written as~:
\bea
\frac{1}{F(p)}&=&\frac{e^2}{i(2\pi)^{d}}\frac{(d-2)\xi}{2p^2}
\left[(p^2+m^2){\cal I}(0,1,1)-{\cal I}(0,0,1)\right],\nn\\
\frac{{\cal M}(p)}{F(p)}&=&\frac{e^2}{i(2\pi)^{d}}m(d-1-\xi)
{\cal I}(0,1,1)\;.
\eea
The longitudinal part of the vertex can be straightforwardly obtained from 
eq.~(\ref{Lvertex}). The transverse 
coefficient are found to be
\bea
\tau_i&=&\frac{e^2}{i(2\pi)^{d}}\Bigg\{t_{i,0}I(1,1,1)+t_{i,1}[-(k\cdot
q){\cal I}(0,1,1)+(p\cdot q){\cal I}(1,0,1)
\nn\\&+&
q^2I(1,1,0)]
+t_{i,2}({\cal I}(0,1,1)+{\cal I}(1,0,1)-2I(1,1,0))
\nn\\&+&
t_{i,3}\left({\cal I}(0,1,1)+
{\cal I}(1,0,1)-2\frac{{\cal I}(0,0,1)}{m^2}\right)\nn\\
&+&t_{i,4}({\cal I}(0,1,1)+{\cal I}(1,0,1))+t_{i,5}\frac{{\cal I}(0,1,1)
-{\cal I}(1,0,1)}
{k^2-p^2}\Bigg\}\;,
\eea           
where the $t_{i,j}$, $j=1\ldots 5$ are basic functions of $k,p$ and $q$, 
listed in \cite{davydychev}.

\end{itemize}

Because of the complexities of expressions,
the non perturbative 
structure of the vertex can hardly be obtained from them
in arbitrary dimensions. 
However, in 3 dimensions, the expressions simplify to an extent that
a less complicated non perturbative structure of 
the transverse part of the fermion-boson interaction can be seen.
In view of the perturbative requirements, we analyse below
some of the \emph{ans\"atze} proposed~:

\begin{itemize}

\item {\bf \underline{Curtis and Pennington~:}}

The \emph{ansatz} of Curtis and Pennington,~\cite{CP1} was explicitly 
constructed to
reproduce the corresponding result in the weak coupling regime in the relevant
kinematical domain when momentum in one fermion leg is much greater
than in the other leg, i.e.,
\bea
\Gamma^\mu_T(k,p)\simeq
\frac{\alpha\xi}{8}\ln\left(\frac{k^2}{p^2} \right)
\left[-\gamma^\mu+\frac{k^\mu \slsh{k}}{k^2} \right]\;.  \label{asymptotic}
\eea

\item {\bf \underline{Dong, Munczek, and Roberts~:}}

This \emph{ansatz} reproduces eq.~(\ref{asymptotic}) 
for 4 dimensions although it 
employs a different set of transverse coefficients as compared to the
CP vertex.

\item {\bf \underline{Bashir and Pennington I~:}}

It also satisfies eq.~(\ref{asymptotic}) in addition to reducing the gauge
dependence of the critical coupling in comparison with the CP-vertex.

\item {\bf \underline{Bashir and Pennington II~:}}

With the appropriate choice of the the constrained functions 
$W_1$ and $W_2$,
defining the
transverse vertex, it should match the perturbative results in the
weak coupling regime.

\item {\bf \underline{Bashir, K{\i}z{\i}lers\"u, and
Pennington~:}}

For the massless case, the constraint obtained from it 
would reproduce the wavefunction
renormalization to all orders with the correct exponent. 

\end{itemize}

For the case of QED3 specifically,

\begin{itemize}

\item {\bf \underline{Curtis and Pennington~:}}

The CP vertex which was designed for QED4 also reproduces correct
perturbative vertex at the one loop in QED3 in the limit when
momentum in one of the fermion legs is much greater than in the other~:
\be
\Gamma^\mu_T(k,p)\simeq
\frac{\alpha\xi}{8} \frac{\pi}{\sqrt{p^2}}
\left[-\gamma^\mu+\frac{k^\mu \slsh{k}}{k^2} \right]\;.
\label{asymptotic3}
\ee

\item {\bf \underline{Burden and Roberts~:}}

The choice of the parameter $a$ which leads to an almost gauge
independent chiral condensate from $\xi=0-1$ has the undesirable feature of
mismatch against perturbative results in the weak coupling regime
even for the limit in eq.~(\ref{asymptotic3}).

\item {\bf \underline{Burden and Tjiang~:}}

The one parameter ($\beta$) family of vertex \emph{ans\"atze} deconstructed
by Burden and Tjiang was done assuming $\beta$  to be gauge
independent.  The perturbative result eq.~(\ref{asymptotic3}) shows that 
this could not be the case.

\item {\bf \underline{Bashir and Raya~:}}

This \emph{ansatz} was constructed to reproduce perturbative results in
the weak coupling regime not only in the above-mentioned kinematical regime, 
but for all momentum regimes.

\end{itemize}

In the construction of non perturbative vertices, a next step could be to
have a vertex in arbitrary dimensions reproducing the one loop vertex
correctly. Going beyond one loop should come after that. Several
advances have been made in solving the integrals which appear in the
two loop calculation of the fermion propagator,~\cite{FJTV},
and the vertex,~\cite{2davy1,2davy2,2davy3,2davy4,glover1}.

\section{LKF Transformation Laws}

 In a gauge field theory, Green functions transform in a specific 
manner under a variation of gauge. In QED
these transformations carry the name Landau-Khalatnikov-Fradkin 
transformations (LKFT),~\cite{LK1,LK2,F1}.
These were derived also by Johnson and Zumino through functional
methods,~\cite{JZ1,Z1}, and by others~\cite{Oku,IBB,Sono}~\footnote{Fukuda, Kubo and Yokoyama have looked for 
a possible formalism where renormalization constants of the wave function 
are in fact gauge invariant~\cite{Fukuda}.}. 
LKFT are nonperturbative in
nature and hence have the potential of playing an important role in 
addressing the problems of gauge invariance which plague the strong 
coupling studies of SDEs. In general, the 
rules governing these transformations are far from simple. The fact 
that they are written in coordinate space adds to their complexity. As 
a result, these transformations have played less significant and
practical role in the study of SDE than desired. 

As compared to the LKFT, WGTI are simpler to use and,
therefore, have been extensively implemented in the SDE
studies which are based either upon gauge technique, e.g., 
\cite{Salam1,SD1,S1,DW1,DW2,D1,Keck1,D2,Hoshino}, 
or upon making an {\em ansatz} for the full fermion-boson 
vertex, e.g., \cite{BC,CP1,BP1,BP2,CP3,CP2,ABGPR1,AGM1}. 
WGTI follow from the 
Becchi-Rouet-Stora-Tyutin 
(BRST) symmetry. One can enlarge these transformations 
by transforming also the gauge parameter $\xi$, \cite{Nielsen,Sibold},
to arrive at modified Ward identities, known as Nielsen identities (NI).
An advantage of the NI over the conventional Ward
identities is that $\partial/\partial \xi$ becomes a part of the
new relations involving Green functions. This fact was
exploited in \cite{Steele,Grassi} to prove 
the gauge independence of some of the quantities related to two-point
Green functions at the one loop level and to all
orders in perturbation theory, respectively. As it is a
difficult task to establish the gauge independence of physical
observables in the study of SDE, NI may play a
significant role in addressing this issue in addition
to Ward identities and LKFT. However, in this section, we
concentrate only on the LKFT.

 The LKFT for the three-point vertex is complicated and
hampers direct extraction of analytical restrictions on its
structure. Burden and Roberts, \cite{BR1}, carried out a numerical 
analysis to compare the self-consistency of various {\em ans\"atze} for
the vertex, 
\cite{BC,CP1,H1}, by means of its LKFT.
In addition to these numerical constraints, indirect analytical
insight can be obtained on the nonperturbative structure of the
vertex by demanding correct gauge covariance properties of the 
fermion propagator. In the context of gauge technique, examples
are \cite{Keck2,Keck3,Waites1}. Concerning the works based upon choosing a
vertex {\em ans\"atze}, references 
 \cite{BKP1,CP1,BP1,BP2,dong,BT1} employ this idea \footnote{A criticism of
the vertex construction in \cite{BT1} was raised in \cite{BKP2}.}. 
However, all the work in the later category has been carried out mostly for
massless QED3 and QED4. The masslessness of the
fermions implies that the fermion propagator can be written only in
terms of one function, the so-called wavefunction renormalization, $F(p)$.
In order to
apply the LKFT, one needs to know a Green function at least
in one particular gauge. This is a formidable task. However, one
can rely on approximations based on perturbation theory. It is
customary to take $F(p)=1$ in the Landau gauge, an
approximation justified by one loop calculation of the massless fermion
propagator in arbitrary dimensions, see for example, \cite{davydychev}.
The LKFT then implies a power law for $F(p)$ in QED4
and a simple trigonometric function in QED3. To improve upon these
results, one can take two paths

\begin{itemize}

\item
 incorporate the information contained in higher orders of
perturbation theory both in the massless and massive case.

\item

 study how the dynamically generated fermion propagator transforms under
the LKFT.

\end{itemize}

As pointed out
in \cite{BKP1}, in QED4, the power law structure of the wavefunction
renormalization remains intact by increasing order of
approximation in perturbation theory although the exponent of course
gets contribution from next to leading logarithms and so on 
\footnote{For the two loop calculation of the fermion propagator,
see for example \cite{Ross}.}. In \cite{BKP1}, constraint was obtained
on the 3-point vertex by considering a power law where the exponent 
of this power law was not restricted only to the one loop fermion 
propagator. In QED3, the two loop fermion propagator was evaluated in
\cite{BKP2,BKP3,adnan1}, where it was explicitly shown that the 
the approximation $F(p)=1$ is only valid up to one loop, thus violating
the {\em transversality condition} advocated in \cite{BT1}. The result
found there was used used in \cite{adnan2} to find the improved
LKF transform. In the following, we try to summarize in relatively more 
detail the progress that has and is being made in this direction.

\subsection{LKFT in QED4}

In massless QED4 the LKFT imply that the fermion propagator is 
multiplicatively renormalizable, i.e., $F(p)=A (p^2)^{\nu}$.  
Brown and Dorey,~\cite{Brown}, 
argued that an arbitrary \emph{ansatz} for the vertex does not satisfy the 
requirement of the multiplicative renormalizability (MR). It was realized that
neither the bare vertex nor the BC-vertex were good enough to fulfill the 
the demands of MR. 

\begin{itemize}

\item  {\bf {\underline{Bare Vertex:}}} 

 For the massless case, the bare vertex leads to the following
SDE for the fermion propagator~:
\bea
  \nn  \frac{1}{F(p)} &=& 1 - \frac{\alpha \xi}{4 \pi \nu} \;
\left[ \frac{2F(p)}{\nu+2} - F(\Lambda) \right]  \;.
\eea
This equation does not have any solution for $A$ and $\nu$, except in
the Landau gauge where $A=1$ and $\nu=0$. Therefore, for the bare vertex,
$ F(p)$ has multiplicatively renormalizable solution only in the
Landau gauge.

\item    {\bf {\underline{Ball and Chiu:}}} 

If instead of the bare vertex, we would have used the BC-vertex, 
an analogous calculation would have led us to the following equation for
$F(p)$ in the massless limit
\bea
\nn
  \frac{1}{F(p)} &=& 1 - \frac{\alpha \xi}{4 \pi \nu} \; F(\Lambda)
\\
\nn
&+&  \frac{3 \alpha}{16 \pi} \; F(p) \; 
\left[ \frac{5}{2} + 2 \pi {\rm cot} \nu -  \frac{1}{\nu} 
- \frac{2}{\nu+1} - \frac{1}{\nu+2} +  {\rm ln} \frac{\Lambda^2}{p^2}   
\right]  \;.
\eea
The explicit presence of the term ${\rm ln} \Lambda^2/p^2$ spoils the
multiplicatively renormalizable solution.

\item  {\bf {\underline{Curtis and Pennington:}}} 

Curtis and Pennington,~\cite{CP1}, realized that the transverse vertex is
crucial in ensuring the multiplicative renormalizability of the fermion
propagator. Usage of their proposal into the SDE for the fermion propagator renders the latter
multiplicatively renormalizable giving
\bea
    F(p) &=& \left[ \frac{p^2}{\Lambda^2}  \right]^{\nu}  \;
\label{RFP1}
\eea
with $\nu=\alpha \xi / 4 \pi$.

\item {\bf {\underline{Bashir and Pennington I:}}} 

This  proposal  
also generates exactly the same fermion propagator as in eq.~(\ref{RFP1}).

\item {\bf {\underline{Bashir and Pennington II:}}} 

The generalization of the above ansatz still ensures Eq.~(\ref{RFP1}).

\item {\bf {\underline{Dong, Munczek and Roberts:}}}

Dong, Munczek and Roberts imposed the so called transversality
condition on the transverse vertex to make sure that the LKFT were
satisfied both in 3 and 4 dimensions simultaneously. 
This vertex, however, exhibits logarithmic kinematic singularities
in 3 dimensions. 

\item {\bf {\underline{Bashir, K{\i}z{\i}lers\"u and Pennington:}}}

 It is easy to see that the 
solution $F(p) = (p^2/\Lambda^2)^{\nu}$, with 
$\nu=\alpha \xi  /4 \pi$  correctly embodies the leading log terms which 
are proportional to 
$\alpha \xi$, $(\alpha \xi)^2$, $(\alpha \xi)^3$, etc. as is obvious in the
expansion  $F(p) = 1 + \nu \; {\rm ln} {p^2}/{\Lambda^2}
+ ({\nu^2}/{2!}) {\rm ln}^2 {p^2}/{\Lambda^2} + \cdots$. Crucially, this is not
the general solution, nor it is in agreement with perturbation theory
beyond the leading logs. The general solution is 
$F(p) = (p^2/\Lambda^2)^{\gamma}$ where $\gamma$ is not zero in the 
Landau gauge. In fact, $\gamma= a \xi - \frac{3}{2} \, 
a^2 + \frac{3}{2} \, a^3 +  {\cal O}(a^4)$, where $a=\alpha/ 4 \pi$
,~\cite{Tarasov}. The most 
general non perturbative construction of the transverse vertex required by the
multiplicative renormalizability of the fermion propagator, replacing the
exponent $\nu$ by $\gamma$ was carried out by Bashir, K{\i}z{\i}lers\"u and Pennington~\cite{BKP1}. 

\end{itemize}

\subsection{LKFT in QED3}

In massless QED3 the LKFT imply that if we take the wavefunction
renormalization to be $1$ in the Landau gauge, then in an arbitrary covariant
gauge~\footnote{The gauge dependence of the infrared behaviour of this theory has also been studied in~\cite{MRS} with a different approach.}  
\begin{eqnarray}
F(p)&=&1-\frac{\alpha\xi}{2 p} \; 
{\rm tan}^{-1} \left[ \frac{2 p}{\alpha\xi} \right] \;. \label{FP3LKF1}
\end{eqnarray}
However, it has been shown that although this 
condition is satisfied at one loop order, it gets violated at the 
two loop order, leading to the following expression for the 
fermion propagator,~\cite{adnan2}:
\begin{eqnarray}
   F(p)&=& 1 - \frac{\pi \, \alpha \xi}{4 p}  + 
               \frac{\alpha^2 \xi^2}{4p^2}
              - \frac{3 \alpha^2}{4 p^2} \left(\frac{7}{3}-\frac{\pi^2}{4} 
              \right) +{\cal O}(\alpha^3) \,.
\end{eqnarray}
Now, starting from the value of this expression in the Landau gauge,
LKFT yields
\bea
 F(p) &=& 1\;  - \; \frac{\alpha \xi}{2p} \, {\rm tan}^{-1} \, 
\frac{2p}{\alpha \xi} \;-\;  
\frac{(28-3 \pi^2)p^2 \alpha^2}{(\alpha^2 \xi^2 + 4p^2)^2} \;. 
\label{FP3LKF2}
\eea
An \emph{ansatz} for the vertex should yield this fermion
propagator through the SDE, guaranteeing its correctness till the two loop
order. Bare vertex or the Ball-Chiu vertex again do not come up to these
expectations. The CP-vertex, BP-vertex and the DMG-vertex reproduce 
eq.~(\ref{FP3LKF1}) correctly but not eq.~(\ref{FP3LKF2}).

\begin{itemize}

\item  {\bf {\underline{Burden and Roberts:}}} 

This vertex satisfies the WGTI but fails to comply with the requirements of the LKFT beyond the lowest order.

\item  {\bf {\underline{Burden and Tjiang:}}} 

The deconstruction of Burden and Tjiang reproduces 
eq.~(\ref{FP3LKF1}) correctly but not eq.~(\ref{FP3LKF2}).

\item  {\bf {\underline{Bashir and Raya:}}}

By construction this vertex 
satisfies the LKFT at the one loop order and yields a fermion
propagator which satisfies its LKFT at the two loop order.

\end{itemize}

\subsection{LKFT and the Massive Fermion Propagator}

      The discussion on the LKFT so far is limited to the cases when in the
massless limit the vertex function results in the correctly behaved fermion 
propagator. To include the discussion of the massive fermions, we need to go 
beyond it. 
In~\cite{BaRa2},  LKF transformed fermion propagator
in massive QED3 and QED4 was obtained \footnote{In the context of 
gauge technique,
gauge covariance of the spectral functions in QED was studied in 
\cite{Keck2,Keck3,Waites1}.}, starting from the 
simplest input which
corresponds to the lowest order of perturbation theory, i.e.,
$S(p)=1/ ({\not \! p}-m_0)$ in the Landau gauge. On LKF transforming, 
the fermion propagator in an arbitrary covariant gauge was obtained.
In the case of QED3, the result can be written in terms of basic functions
of momenta. In QED4, the final expression is in the form of
hypergeometric functions where
coupling $\alpha$ enters as parameter of this transcendental
function. 
Results are compared with the one loop expansion of the fermion
propagator in QED4 and QED3, \cite{Reenders,BashirR1}, and perfect
agreement is found up to terms independent of the gauge parameter at one loop,
a difference permitted by the structure of the LKFT\footnote{The 
references for the mathematical details behind this exercise can be found 
in~\cite{Moch,Zsums1,Zsums2,tables,moretables}}.
In QED3, the results are~:
\bea
F(p)&=&-\frac{\alpha\xi}{2p}
\arctan{\left[\frac{2p}{(2m_0+\alpha\xi)} \right]}
+ \frac{2p(4p^2+\alpha^2\xi^2)}{ \phi  }  
\nn\\ &&
-\frac{\alpha\xi (4p^2+\alpha\xi
(2m_0+\alpha\xi))}{\phi}
\arctan{\left[\frac{2p}{(2m_0+\alpha\xi)}\right]}\;,  \label{LKFF3} \\
{\cal M}(p)&=&\frac{8p^3 m_0}{\phi } \;,   \label{LKFM3}
\eea
where
\bea
\nn
\phi=2p(4p^2+\alpha\xi(2m_0+\alpha\xi))-
\alpha\xi(4p^2+(2m_0+\alpha\xi)^2)\arctan{\left[\frac{2p}{(2m_0+\alpha\xi)}
\right]}.
\eea
In QED4, the corresponding expressions are
\bea
\nonumber \\ F(p)&=&
\frac{ \Gamma(1+\nu)}{
2 m_0^2 \; \Gamma(3+\nu) \;  {~}_2F_1\left(1+\nu,3+\nu;3;
-\frac{p^2}{m_0^2} \right)} \; \left( \frac{m_0^2}{\Lambda^2} 
\right)^{-\nu} \nonumber  \\
&&
\times\Bigg[4m_0^2\Gamma^2(2+\nu) \; {~}_2F_1^2\left(1+\nu,2+\nu;2;
-\frac{p^2}{m_0^2} \right)
\nn\\&&
+p^2 \; \Gamma^2(3+\nu) \;
_2F_1^2\left( 1+\nu,3+\nu;3;-\frac{p^2}{m_0^2}\right) \Bigg] \;,
\label{ourresultF}  \\ \nonumber \\
{\cal M}(p)&=&\frac{2m_0 \; {~}_2F_1\left(
1+\nu,2+\nu;2;- \frac{p^2}{m_0^2} \right)}
{(2+\nu) \; {~}_2F_1\left(1+\nu,3+\nu;3;- \frac{p^2}{m_0^2} \right)} \;,
\label{ourresultM}
\eea
with $\nu=\alpha\xi/(4\pi)$. These results match onto one loop 
perturbative results upto gauge parameter independent term as expected.
In fact, these results should correctly reproduce all order terms in
perturbation theory, of the type $(\alpha \xi)^n$, as pointed out 
in~\cite{BD2}. For the massless case, these results reduce to the well-known
expressions, including the power-law for the wavefunction renormalization 
function, as expected.

\subsection{LKFT and DCSB}

As LKFT are non perturbative in nature, they would not only tell us
how a perturbative Green function will transform with the variation of 
gauge but also how a non perturbative one would do so. The first
exercise to our knowledge in this connection for the dynamically generated 
mass function has been carried out in~\cite{BR4} in the context of QED3.
Thus knowing the solution of the SDE for the mass function in the Landau
gauge, one can LKFT to any other gauge. One can parametrize the
solution in the Landau gauge for practically any choice of the 
full vertex  approximation with the following expressions\footnote{In
this section, we shall use the notation $F(p,\xi)$ and $\M(p,\xi)$ for 
the wavefunction renormalization and the mass function in the covariant 
gauge $\xi$.}~:
\bea
F(p,0) &=& F_{0} \theta(m_{0}-p) +\theta(p-m_{0})  \label{fullparam} \;,
\\
{\cal M}(p;0) &=& M_{0} \left[ \theta(m_{0}-p) + \frac{m_{0}^2}{p^2} \theta(p-m_{0}) 
\right]
\;, \label{param1}
\eea
The wavefunction renormalization behaves like a
constant (different from unity) for low momentum, and tends to
one as $p\to\infty$.  For the rainbow approximation $F_{0}=1$.
The mass function is a constant for low momentum and falls off like
$1/p^2$ for large momentum. Now we carry out the LKFT exercise.
Consequently, in the large-$p$ limit it is found that
\bea
{\cal M}(p;\xi)&=&  
 \frac{C_{3}(\xi)}{p^2}   +{\cal O}\left( \frac{1}{p^3}\right) \;,\\
F(p;\xi) &=&1+{\cal O}\left( \frac{1}{p}\right) \;,
\eea
where
\bea
C_{3}= m_{0}^2 M_{0} + \frac{4 a m_{0} (m_{0}^2 F_{0} + 
3 M_{0}^2)}{3 M_{0} \pi} \;,  \nn
\eea
with $a=\alpha \xi/2$ and in the low-$p$ domain
\bea
{\cal M}(p;\xi)&=& \frac{C_{1}(\xi)}{C_{2}(\xi)} + {\cal O}(p^2) \;, \\
F(p;\xi)&=&-\frac{C_{1}^2(\xi)}{C_{2}(\xi)}-C_{2}(\xi) p^2  
+ {\cal O}(p^4)\;,
\eea
where
\bea
C_{1}(\xi)&=&\frac{m_{0} M_{0}}{a^4}
\left( 3m_{0}-\frac{4a}{\pi}\right)
+ \frac{2am_{0}}{\pi
(a^2+m_{0}^2)}\left( \frac{F_{0}}{M_{0}}-
\frac{m_{0}^2M_{0}}{a^4}\right)
\nn\\&&
-\frac{2}{\pi}\left(\frac{F_{0}}{M_{0}}
+\frac{3m_{0}^2M_{0}}{a^4}\right)
\arctan{\frac{m_{0}}{a}}\nn\\
C_{2}(\xi)&=&\frac{2}{3M_{0}^2\pi}\left[ 
\frac{am_{0}(3a^2 F_{0}+5m_{0}^2 
F_{0}-2M_{0}^2)}
{(a^2+m_{0}^2)^2}
-3 F_{0} \arctan{\frac{m_{0}}{a}}\right]. \nn
\eea
One concludes from these expressions that
qualitative features of the mass function and the wavefunction renormalization
remain intact in any other gauge, i.e., both the functions behave in an
identical fashion in the large and low-momentum regions as they do in the
Landau gauge.

As the SDE of higher point Green functions are intrincately related to it,
a meaningful truncation scheme which maintains the key features of
the theory, such as gauge invariance, continues to be a challenging
problem. Along with other
guiding principles, a correct inclusion of the  LKFT of the fermion 
propagator and the 
fermion-boson interaction is crucial for arriving at reliable conclusions.
The above provides a preliminary study in this connection. 
Encouragingly, one finds expected behaviour of the fermion propagator for the
asymptotic regimes of momenta for all  $\xi$, a fact supported by earlier
direct numerical solutions of the SDE (where no reference is made
to the LKFT) in a small region of $\xi$ close
to the Landau gauge~\cite{BHR,BR1}.
The initial results for quenched QED3 presented here provide the starting 
point for a more rigorous and exact numerical study. In a recent work, such an exersice has been carried out~\cite{LKFmom}. There, the gauge dependence of the chiral condensate both in the quenched and unquenched versions of QED3 has been studied.

The need for such an exercise becomes all the more essential due to the findings reported in~\cite{FRDM} for the chiral condensate. Although they employ a truncation scheme which preserves Ward identities, gauge dependence of the condensate is conspicuous specially for the unquenched case even in the immediate vicinity of the Landau gauge. It was pointed out in~\cite{LKFmom}, that the origin of this unwanted behaviour owes itself to the fact that LKFTs were not satisfied, an essencial consequence of gauge invariance. Once these are incorporated, one obtains a practically gauge independent value of the chiral condensate for a very broad range of values of the covariant gauge parameer, Following is a representative figure for quenched QED3 to demonstrate the miraculous role played by the LKFT.

\begin{center}
\epsfig{file=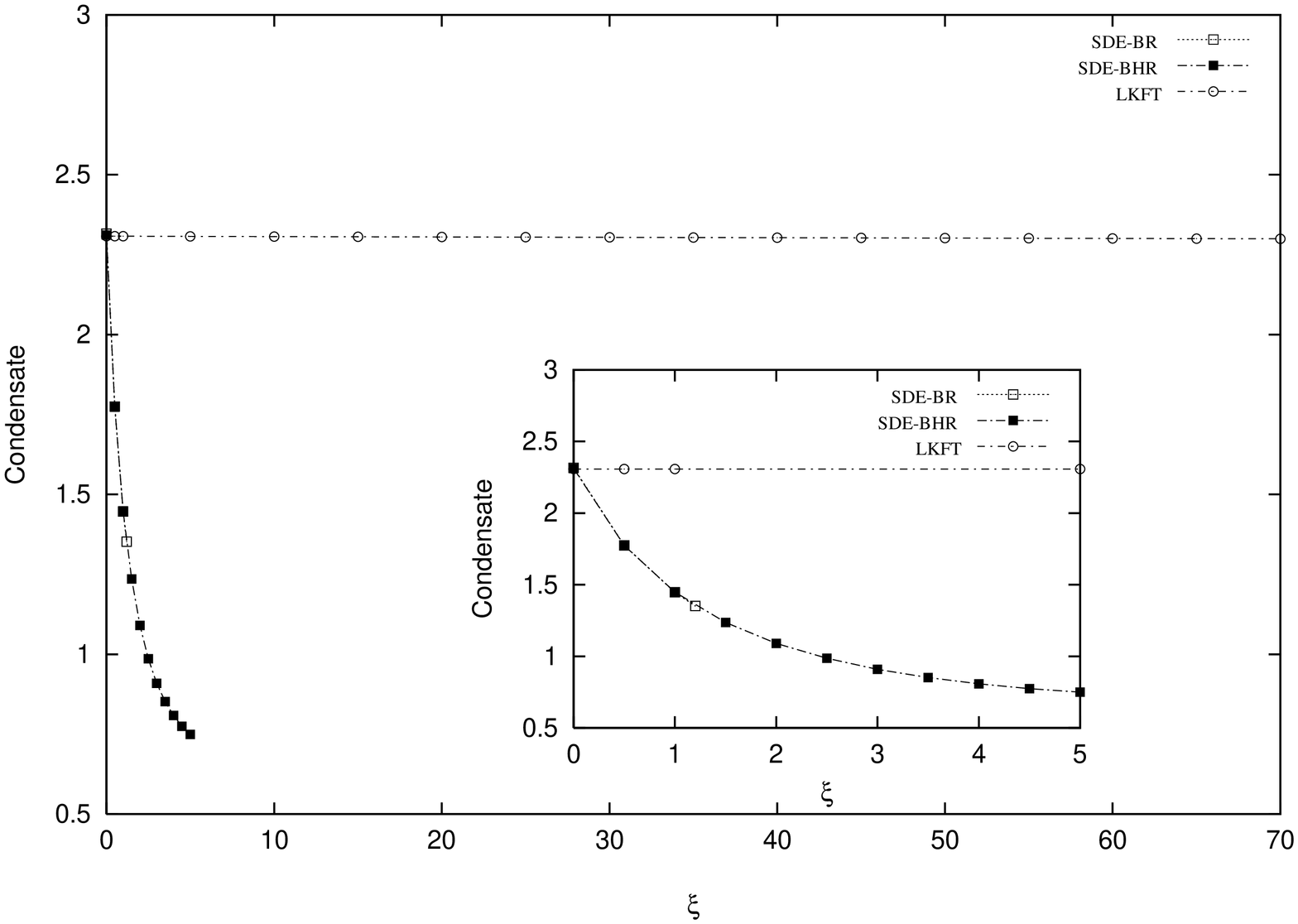,width=0.4\textwidth,angle=-90}
\end{center}
\begin{center}
{\sl Figure 16. Gauge dependence of the chiral condensate in quenched QED3 employing the
bare vertex.}
\end{center}

One may wonder why we worry about the WGTI, kinematic singularitiesm CPT Symmetries and Perturbation Theory if, as suggested in~\cite{LKFmom}, the LKFT are apparently sufficient to obtain the objective. The answer is that all the ingredientes are crucial in the hunt for a more reliable result in one gauge, i. e., the Landau gauge. This is where the choice of the full vertex plays an important role and the closer we are to the true vertex, the better it is. Once we know the result in the Landau gauge, LKFT will guide us along the path of varying the gauge. One can then go
on to take up the more interesting case of QED4. The solution to this problem
will in turn be the starting point for a study of QCD where the non-abelian 
nature of interactions, so essential for both confinement and asymptotic
freedom, will further complicate the problem.

\section{Conclusions}

The non perturbative knowledge of the Green functions has vital importance
in QCD when the coupling is not small enough to carry out 
a perturbative expansion and also in the dynamical models beyond the 
standard model of particle physics, such as the technicolour models, top condensate models, 
top color models etc. Schwinger-Dyson equations provide a natural tool for
the continuum studies of all these scenarios. However, any non perturbative
truncation scheme results in the loss of many deservedly sacred properties
of these quantum field theories, including the gauge covariance of these
Green functions resulting in the gauge dependent physical observables!
Because of the highly involved structure of the Schwinger-Dyson equations,
it is a difficult task to look for a truncation scheme which
can solve this problem. However, a lot of progress is being made in simpler
theories such as quenched QED in 3 and 4 dimensions to restore the gauge 
identities and render the physical observables associated with the fermion propagator
gauge independent. We have summarised the main attempts in this direction.
Based upon these efforts, we hope that unquenching these theories and 
studying non-abelian theories such as QCD would become relatively less 
daunting and more accessible.

\noindent
{\bf Acknowledgements~:} AR wishes to thank the invitation for this 
contribution and to the Alvarez-Buylla grant under project 292/04.
AB acknowledges the CIC and CONACyT grants 4.10 and 46614-I respectively.

\end{document}